\documentclass{article}
\usepackage[utf8]{inputenc} 
\usepackage[T1]{fontenc} 
\usepackage[margin=28mm, bottom=30mm]{geometry}

\usepackage{mathrsfs}

\usepackage{physics}
\usepackage{amsfonts}
\usepackage{mathtools} 

\usepackage{graphicx} 
\usepackage{float} 
\usepackage{subfig} 

\usepackage{tabularx} 
\usepackage[hidelinks]{hyperref} 
\usepackage{xcolor} 

\usepackage{authblk} 

\usepackage[backend=biber, maxnames=10, uniquelist=false, sorting=none]{biblatex}
\usepackage[textwidth=22mm]{todonotes}

\DeclareMathOperator{\Ricci}{Ric}
\DeclareMathOperator{\Var}{Var}

\newcolumntype{P}[1]{>{\centering\arraybackslash}p{#1}}

\newcommand{\mycomment}[1]{}

\usepackage{stackengine}

\begin{document}

\title{Scalar curvature for metric spaces:\\
Defining curvature for Quantum Gravity without coordinates}

\author{Agustín Silva\footnote{Contact: agustin.silva@ru.nl} }
\author{Jesse van der Duin\footnote{Contact: jesse.vanderduin@ru.nl}}
\affil{$^{*\dag}\!$High Energy Physics Department, Institute for Mathematics, Astrophysics, and Particle Physics, Radboud University, Nijmegen, The Netherlands.}

\date{\today}

\maketitle

\begin{abstract}
Geometrical properties of spacetime are difficult to study in non-perturbative approaches to quantum gravity like Causal Dynamical Triangulations (CDT), where one uses simplicial manifolds to define the gravitational path integral, instead of Riemannian manifolds.
In particular, in CDT one only relies on two mathematical tools, a distance measure and a volume measure.
In this paper, we define a notion of scalar curvature, for metric spaces endowed with a volume measure or a random walk, without assuming nor using notions of tensor calculus.
Furthermore, we directly define the Ricci scalar, without the need of defining and computing the Riemann or the Ricci tensor a priori.
For this, we make use of quantities, like the surface of a geodesic sphere, or the return probability of scalar diffusion processes, that can be computed in these metric spaces, as in a Riemannian manifold, where they receive scalar curvature contributions.
Our definitions recover the classical results of scalar curvature when the sets are Riemannian manifolds.
We propose two methods to compute the scalar curvature in these spaces, and we compare their features in natural implementations in discrete spaces.
The defined generalized scalar curvatures are easily implemented on discrete spaces, like graphs.
We present the results of our definitions on random triangulations of a 2D sphere and plane.
Additionally, we show the results of our generalized scalar curvatures on the quantum geometries of 2D CDT, where we find that all our definitions indicate a flat ground state of the gravitational path integral.
\end{abstract}

\section{Introduction}\label{sec:intro}
Imagine the situation where someone puts you in a space where you are only given a ruler to measure distances in that space, and a way of measuring the volume of a region in that space.
You are allowed to walk wherever you want inside the space, but not outside.
Then, whoever put you in that space, commands you to determine if the space you are in has intrinsic curvature or not. You do not have any extra structure, like light rays bent by a near curvature source, nor any matter fields that you can use.
This would be akin to living in a simplicial manifold or even a graph.
What would you do?
In this paper, we define notions of scalar curvature for these kinds of situations.
More formally, we define coarse-grained notions of scalar curvature in metric spaces endowed with a volume measure or a random walk. In particular, our measures are best suited for discrete spaces, like simplicial manifolds, or graphs.
But before we dive into these definitions, we address the question: why one should be interested in defining curvature in these spaces? 


The long search for a theory of quantum gravity has provided us with a large variety of ideas and different approaches to join general relativity with quantum mechanics \cite{loll2022quantum}.
Among those approaches is Causal Dynamical Triangulations (CDT), a non-perturbative approach to quantize gravity.
In a nutshell, CDT is a lattice approach to study the underlying quantum theory of gravity.
Analogue to lattice QCD, it uses discrete quantum field theory to study the gravitational path integral \cite{Loll2019}.
It discretizes spacetime using simplicial building blocks that respect causality, distinguishing between time- and space-like distances in its construction.
In this approach the gravitational path integral is given by the continuum limit of the sum over all possible simplicial manifolds without the need for coordinates, making the theory manifestly diffeomorphism invariant.

The notion of curvature is fundamental to the discussion of spacetime in the framework of General Relativity.
In order to study the geometry of the spacetimes appearing in non-perturbative quantum gravity, one can expect curvature to be an important quantity to understand \cite{loll2023curv}. 
However, studying curvature in the discrete setting of simplicial manifolds is challenging, because the metric of the simplicial manifolds is not smooth.
To be able to study curvature in discrete approaches to quantum gravity like CDT, different methods have to be considered. 
In this work we introduce generalizations of the Ricci scalar curvature based on the volume of spheres and the return probability of a random walker, that are especially suited for numerical computation.
These generalized definitions of scalar curvature are motivated by their use in CDT, but can also be applied to other discrete settings like a graph.



Previously, several notions of generalized curvature in simplicial manifolds have been proposed.
A standard notion of scalar curvature in simplicial manifolds was introduced with Regge calculus \cite{Regge1961}, which measures scalar curvature using the deficit angle of the simplicial building blocks.
When applying this prescription to 4D CDT, the continuum limit of this notion of curvature is not defined, because the curvature is given at the scale of the UV cut-off, the lattice discretization length of spacetime. Therefore, for this framework one has to study curvature using a different prescription.

Another generalized notion of curvature is the prescription of Ollivier \cite{Ollivier2009}.
Here the Ricci curvature tensor is defined by using the transport distance between two geodesic spheres.
In this paper, we take Ollivier's idea of defining curvature indirectly, not from taking derivatives of the metric tensor, but instead, by computing well-defined quantities in metric measure spaces, that depend on the Ricci scalar in a Riemannian manifold.
In \cite{Klitgaard2018} an adaption of Ollivier's idea is made with the purpose of applying it to CDT, and is successfully applied, obtaining promising results.
They do this by making a generalization of the Ricci tensor, called by the authors the Quantum Ricci Curvature (QRC), which is proportional to the Ricci curvature when applied to a Riemannian manifold.
The QRC offers a computationally feasible, but expensive calculation in metric measure spaces; it requires the computation of the distance between all pairs of points of two geodesic spheres.

The idea behind the notions of curvature presented in \cite{Ollivier2009} and \cite{Klitgaard2018} is to compute a quantity that depends on the Ricci curvature in a Riemannian manifold, and use this quantity to define curvature in a more general setting.
We would like to emphasize the well previously understood fact, that these are particular cases of the general idea of: instead of defining curvature directly, by taking derivatives of the metric tensor, it is possible to define curvature indirectly, by computing quantities that depend on the curvature in a Riemannian manifold, that only rely on a distance function and some additional structure, like a volume measure or a diffusion process. This indirect/implicit definition of curvature is the method we will use in the following to establish some new ways to determine the scalar curvature of metric spaces.

Our method to define scalar curvature is similar in nature to the QRC, so, it is worth mentioning some of its features.
More specifically, the QRC is defined through the average sphere distance $\bar{d}\left(S_{p}^{\delta}, S_{p'}^{\delta}\right)$, between a pair of spheres with radius $\delta$, centred at points $p$ and $p'$, which are themselves separated by geodesic distance $\delta$. The average sphere distance $\bar{d}\left(S_{p}^{\delta}, S_{p'}^{\delta}\right)$ is the average distance between pairs of points, one in each sphere \cite{Klitgaard2018}. 
\mycomment{That is, on a $d$-dimensional manifold:
\begin{equation}
	\bar{d}\left(S_{p}^{\delta}, S_{p'}^{\delta}\right)
	:= \frac{1}{\norm{S_{p}^{\delta}}} \frac{1}{\norm{S_{p'}^{\delta}}}
	\int_{S_{p}^{\delta}} \dd[D-1]{q} \sqrt{h} \int_{S_{p'}^{\epsilon}} \dd[D-1]{q'} \sqrt{h'} \, d\qty(q, q'),
\end{equation}
where $S_p^\delta$ denotes the geodesic sphere at $p$ with radius $\delta$; $h, h'$ are the determinants of the induced metrics on the spheres; $\norm{S_p^\delta}$ denotes the volume of the sphere with respect to the metric, i.e. $\norm{ S_p^\delta } = \int_{S_p^\delta} \dd[D-1]{q} \sqrt{h}$; and $d(q, q')$ denotes the geodesic distance between $q$ and $q'$.
Now, in a $D$-dimensional Riemannian manifold the average sphere distance has an expansion for small $\delta$ of the form
\begin{equation}
	\frac{ \bar{d} \qty( S_{p}^{\delta}, S_{p'}^{\delta} )}{\delta}
	= C_{1D} - \delta^2 \qty\big(C_{2D} \Ricci\qty(v, v) + C_{3D} R) + \order{\delta^4},
\end{equation}
where $C_{iD}$ are dimensionally dependent positive constants, $\Ricci(v, v)$ is the Ricci tensor in the direction $v$, joining $p$ and $p^{\prime}$, and $R$ the Ricci scalar at $p$.}

In general, the outcome of the calculation of the average sphere distance is normalized by the scale $\delta$, and is parameterized as
\begin{equation}\label{eq:curv-profile-qrc}
	\frac{ \bar{d}\left(S_{p}^{\delta}, S_{p^{\prime}}^{\delta}\right)}{\delta}\coloneq C \qty\big(1 - K\qty(p, p^{\prime}, \delta)).
\end{equation}
Here the quantity $K(p, p', \delta)$ defines the \emph{quantum Ricci curvature}, and it captures any deviation from constancy as a function of $\delta$.
The prefactor $C$ is a positive real value, defined by $C \coloneq \lim_{\delta \rightarrow 0} \bar{d}\left(S_{p}^{\delta}, S_{p'}^{\delta}\right) / \delta$.
This prefactor can be shown to be point independent when considering $\bar{d}\left(S_{p}^{\delta}, S_{p'}^{\delta}\right)$ on a smooth Riemannian manifold, only depending on the dimension of the manifold.
However, when measured on a lattice, the limit of the definition has to be substituted with an appropriate construction on the lattice (see \cite{Klitgaard2018, BrunekreefCurvprofile2021} for details).
On a regular lattice the prefactor $C$ has been found to be a constant depending on the choice of lattice \cite{Klitgaard2018}.
On a general lattice $C$ could in principle depend on the points $p$ and $p'$, but previous investigations \cite{Klitgaard2018, Klitgaard2018Quantizing} have demonstrated that some meaningful results can be obtained in more general lattices assuming $C$ to be point independent.
The QRC $K(p, p', \delta)$ defines a notion of (coarse-grained) curvature, including directional information.
It can be shown that in a Riemannian manifold it can be expanded in powers of curvature invariants, containing at the lowest order the Ricci tensor in the direction joining the two points $p, p'$ \cite{Klitgaard2018}. The point-pair $p, p'$ plays the role of the vector indicating this direction.
Any deviation from $C$ in \eqref{eq:curv-profile-qrc} can be interpreted as the existence of non-zero curvature of the space. Since it is out of the scope of this work to fully introduce the QRC, we refer the reader to the recent review \cite{2023arXiv230613782L}.

Crucially, this way of capturing curvature has the disadvantage that the prefactor $C$ is in general unknown.
This is specifically and issue in cases where analytical calculations of the QRC are out of reach, and one has to rely on numerical methods.
To determine the value of $C$ in the QRC, one takes the smallest $\delta$ values accessible in the measurements, in CDT this is typically around $\delta=5$, where one expects lattice artefacts to disappear for $5 \leq \delta$.
But, if there are other agents influencing the numerical result, like discretization artefacts or some other noise, one would never know if the resulting number is close or far away from the actual value of $C$, if there is any, since these artefacts usually dominate the small $\delta$ region, and therefore it is hard to be conclusive in the results.
Furthermore, there has been no clean way of disentangling the prefactor $C$ from the quantum Ricci curvature $K$, at least in CDT applications, so one only has access to the average sphere distance measurements.

The motivation for this discussion are correlation functions, which are fundamental objects of quantum field theory.
As such, it is interesting to study these for curvature in the case of quantum gravity.

Using the average sphere distance \eqref{eq:curv-profile-qrc} to compute the correlation functions has the problem that the correlations of the prefactor $C$ are in general mixed with the correlations of the QRC, and make more difficult the interpretation of the correlations.
We will later explain how our definitions of curvature somehow bypass this problem by eliminating the presence of any point dependent multiplicative prefactor, providing definitions of curvature where correlations of the obtained quantities might be easier to interpret than those of using \eqref{eq:curv-profile-qrc}.

The structure of this paper is the following.
In section \ref{sec:implicit-curvature} we construct two methods to define generalized notions of scalar curvature from scratch: one based on sphere volumes and one based on return probabilities.
In section \ref{sec:gsc} we generalize this construction of a scalar curvature definition, and show how these methodologies still work for more general settings.
In section \ref{sec:theory-impl} we implement our definitions for triangulations of the sphere and the plane and discuss the applicability of the different methods to extract curvature.
In section \ref{sec:qsc} we apply our generalized scalar curvatures to the quantum geometries of 2D CDT.
Finally, in section \ref{sec:conclusion} we present our conclusions and discuss them.

\section{Generalized scalar curvatures}\label{sec:implicit-curvature}
We begin with an introduction to our generalized scalar curvatures by explicitly constructing them from scratch for two different types of spaces: metric measure spaces, and metric spaces with a random walk.

To this end, we start by introducing two scale-dependent quantities, which can be computed in a metric space, endowed with a measure or a random walk.
The important property is that these quantities give the Ricci scalar curvature at the lowest order of a curvature expansion, if computed in a Riemannian manifold. We will call these quantities Generalized Scalar Curvatures.

First we give an example for a metric space endowed with a volume measure.
Recall, a Metric Measure Space (MMS) \cite{Sturm2006}, denoted by $(\mathcal{X},d,\mu)$, is a mathematical space $\mathcal{X}$ that has a distance function $d$ and a measure $\mu$.
A sphere is defined as $\mathcal{S}_{p}^{r} = \qty{p' \in \mathcal{X} \,:\, d(p, p') = r}$, with its volume given by $\norm{\mathcal{S}_{p}^{r}} \doteq \mu(\mathcal{S}_{p}^{r})$.
Consider the case where $\mathcal{X}$ is a $D$-dimensional Riemannian manifold, $d$ is the geodesic distance function, and $\mu$ is the invariant integration measure.
Then, $\norm{\mathcal{S}_{p}^{r}}$ has the expansion in $r$ \cite{gray1974volume}
\begin{equation}
    \norm{\mathcal{S}_{p}^{r}}
    = \frac{D\,\pi^{\frac{D}{2}}}{\Gamma\qty(\frac{D +2}{2})} r^{D - 1}
    \qty(1-\frac{\mathcal{R}_{p}}{6D}r^{2} + \order{r^4}),
\end{equation}
where $\mathcal{R}_p$ is the Ricci scalar\footnote{We denote the Ricci scalar with $\mathcal{R}$ instead of the commonly used $R$ to avoid confusion with the sphere radius $R$, which will be frequently used later.} at the point $p$.
This expansion can be used to define the scalar curvature as the first scale-dependent correction.
To construct general definitions of curvature, we manipulate the sphere volume, and define the following (scale-dependent) Generalized Scalar Curvatures ($\mathcal{SC}$)\footnotemark{} 
\footnotetext{We will differentiate between the different methods to extract the scalar curvature with subscripts $\qty{1, 2}$. We will use $\mathcal{SC}_i$ to refer to either of them.}

\begin{equation}
   \mathcal{SC}_{1}\qty(\norm{\mathcal{S}_{p}^{r}})
   \coloneq r \pdv{r}\qty(\frac{\partial\log{\norm{\mathcal{S}_{p}^{r}}}}{\partial\log{r}}) \,\,\,\,\,\,\,\,\,\,\,\, \mathcal{SC}_{2}\qty(\norm{\mathcal{S}_{p}^{r}})
   \coloneq \frac{1}{(\int_{0}^{r} dx \,x\,)}\int_{0}^{r} dx \,x\, \left ( \frac{\partial\log{\norm{\mathcal{S}_{p}^{r}}}}{\partial\log{r}} -\frac{\partial\log{\norm{\mathcal{S}_{p}^{x}}}}{\partial\log{x}} \right).
   \label{eq:sc1S}
\end{equation}

In a Riemannian manifold, these two quantities can be expanded at small distance scales, obtaining
\begin{equation}
    \mathcal{SC}_{1}\qty(\norm{\mathcal{S}_{p}^{r}})
    = -\frac{2 \mathcal{R}_{p}}{3 D} r^{2} + \order{r^4} \,\,\,\,\,\,\,\,\,\,\,\, \mathcal{SC}_{2}\qty(\norm{\mathcal{S}_{p}^{r}})
    = -\frac{ \mathcal{R}_{p}}{6 D} r^{2} + \order{r^4}.
    \label{eq:expscSphere}
\end{equation}
Notice that they recover the Ricci scalar in the limit $r \rightarrow 0$, up to a factor $-\frac{2}{3D}r^2$ and $-\frac{1}{6D}r^2$ respectively. This is why we call $\mathcal{SC}_{1}\qty(\norm{\mathcal{S}_{p}^{r}})$ and $\mathcal{SC}_{2}\qty(\norm{\mathcal{S}_{p}^{r}})$ generalized scalar curvatures.

Using these manipulations, we can generalize the notion of scalar curvature to any metric measure space, and $\mathcal{SC}_{1}(\norm{\mathcal{S}_p^r})$ or $\mathcal{SC}_{2}(\norm{\mathcal{S}_p^r})$ can be used as a measure of the coarse-grained scalar curvature at that point.
Of course, for large values of $r$ the interpretation of the coarse-grained curvatures $\mathcal{SC}_{1}(\norm{\mathcal{S}_{p}^{r}})$ or $\mathcal{SC}_{2}(\norm{\mathcal{S}_{p}^{r}})$ can be more difficult due to possible higher-order curvature corrections.
However, in an appropriate short-scale range the interpretation is straightforward.
Note that, in cases where $r$ is a discrete variable and/or $\norm{\mathcal{S}_{p}^{r}}$ is not defined for every continuum value of $r$, an analytic continuation is understood in the definition of $\mathcal{SC}_{1}(||\mathcal{S}_{p}^{r}||)$.
Discrete implementations of the derivatives and integrals are needed in explicit calculations.
We discuss such implementations for the discrete examples we consider in the following sections.

Let us use the previous technique to define curvature from a different perspective. One can use similar methods to obtain a scalar curvature definition using the return probability density of a diffusion process. Diffusion processes can be constructed from Markov chains, where the probability of the next event in the chain depends only on the current event of the chain. More specifically, one can model a diffusion process on a metric space by defining a Random Walk in it \cite{10.1007/978-3-662-43920-3_6}. The probability density of the random walker to move from point to point, is equivalent to the probability density of a particle that is being diffused on the space itself \cite{lawler2010random}.

 More formally, a Metric Space with a Random Walk (RWMS), denoted by $\qty(\mathcal{X}, d, \qty{X_{p}} )$, is a mathematical space $\mathcal{X}$ that has a distance function $d$ and a set of random variables  $\qty{X_{p}}$ for each point $p$ in $\mathcal{X}$, where each random variable $X_{p}$ has to be integrable, and have a finite first moment \cite{Ollivier2009, ibe2013elements}.
 
 Since our main goal is to apply our techniques to discrete spaces, we will restrict our discussion to discrete Random Walks. In this scenario, each $X_{p}$ is a random variable describing the displacement from point $p$ to the next point in the random walk. For example, in a square 2D lattice of unit edge length, and for a random walker moving only to nearest neighbours of the lattice, if $p=\{0,0\}$, then $X_{p} \in \{ \{1,0\}, \, \{0,1\}, \, \{-1,0\}, \, \{0,-1\} \}$. If the random walk is non-biased, then the probability distribution of $X_p$ is uniform over the points $\{ \{1,0\}, \, \{0,1\}, \, \{-1,0\}, \, \{0,-1\} \}$. 

Going back to a general discrete case, we define the ``walking time'' $n$, as the number of steps taken for the random walker since its starting point. In this setting, we will denote the probability of the random walker to go from $p$ to $p'$ as $P_{p,p'}^n$, and the probability that the walker returns to $p$ after time $n$, called the return probability, simply as $P_{p}^n=P_{p,p}^n$.

We would like to relate this to continuum manifolds, in order to define some notion of curvature. A formal identification of $P_{p,p'}^n$ with a discrete version of the Heat Kernel $K(p,p' ; \tau)$ is given in \cite{lawler2010random} (see Chapter 1). For this identification to hold,  relation between the number of steps of the random walker $n$ and the diffusion time of the Heat Kernel $\tau$ has to be linear, i.e. $n \propto \tau $. The proportionality constant has to be determined in each setting, and will be of importance in this work.

In particular, one can identify the return probability $P_{p}^n$ with the diagonal part of the Heat Kernel, by doing 
\begin{equation}
    P_{p}^n = K(p,p;\tau).
    \label{eq:identRetPHK}
\end{equation}
This identification has been widely used to study discrete spaces with unknown properties, such as their dimension, see \cite{ambjorn2005spectral, benedetti2009spectral, reitz2023generalised, brunekreef2022phase, durhuus2007spectral, eichhorn2014spectral} for details and examples. 

We use this correspondence between a random walk and a diffusion process to construct a definition of scalar curvature in a RWMS.
Note that we will not assume that every metric space with a random walk satisfies this identity, and we only use it to show that our calculations recover the Ricci scalar in the case of a Riemannian manifold, with the previous diffusion conditions satisfied. 

For this construction, note that in a D-dimensional Riemannian manifold, the heat kernel $K(p,p;\tau)$ has the well-known expansion \cite{vassilevich2003heat}
\begin{equation}
    K(p, p; \tau) = \frac{1}{(4 \pi \tau)^{\frac{D}{2}}} \qty(1+ \frac{\mathcal{R}_{p}}{6}\tau + \order{\tau^{2}}).
\end{equation}
To this expression, we apply several operations to extract the scalar curvature, analogous to the construction used for the sphere volume.
With this in mind, we define the Generalized Scalar Curvatures in a RWMS as
 \begin{equation}
    \mathcal{SC}_{1}\qty(P_{p}^n)
   \coloneq n \pdv{n}\qty(\frac{\partial\log{P_{p}^n}}{\partial\log{n}}) \qquad \mathcal{SC}_{2}\qty(P_{p}^n)
   \coloneq \frac{1}{(\int_{0}^{n} dx \,x\,)}\int_{0}^{n} dx \,x\, \left ( \frac{\partial\log{P_{p}^n}}{\partial\log{n}} -\frac{\partial\log{P_{p}^x}}{\partial\log{x}} \right).
    \label{eq:sc1P}
\end{equation}
In the cases where $n$ is a discrete variable, one is to understand this definition as an analytical continuation in $n$.
We will return to the discretization of these definitions in the following sections. 
Notice that on a Riemannian manifold, and using \eqref{eq:identRetPHK}, \eqref{eq:sc1P} takes the form
\begin{equation}
    \mathcal{SC}_{1}(P_{p}^n)
    = \frac{\mathcal{R}_{p}}{6}\tau + \order{\tau^{2}} \qquad \mathcal{SC}_{2}(P_{p}^n)
    = \frac{\mathcal{R}_{p}}{18}\tau + \order{\tau^{2}}
    \label{eq:expscRetP}
\end{equation}
So we see that $\mathcal{SC}_{1}(P_{p}^n)$ and $\mathcal{SC}_{2}(P_{p}^n)$ recover the Ricci scalar in the limit of $n \to 0$, up to a factor $\frac{\tau}{6}$ and $\frac{\tau}{18}$ respectively.

\vspace{1em}
\noindent
Summarizing, the defined coarse-grained scalar curvatures $\mathcal{SC}_{1}(P_{p}^n)$, $\mathcal{SC}_{1}(||\mathcal{S}_{p}^{r}||)$, $\mathcal{SC}_{2}(P_{p}^n)$ and $\mathcal{SC}_{2}(||\mathcal{S}_{p}^{r}||)$ recover the Ricci scalar for a Riemannian manifold at small scales.
In a general metric space we can use these quantities to define a scale-dependent curvature scalar. Notice that they are not fundamentally different, but just different methodologies to compute the same curvature scalar in the limit of the scales going to $0$.
For example, these scalar curvatures can be used in a graph where one uses the link distance as distance function, and the number of nodes in a set as the volume measure.
In this space, sphere volumes can be measured with discrete radius $r$ and can be used to compute the scalar curvature at every point.
Additionally, a random walk can be implemented to determine scalar curvature equivalently but based on a different computation.
An advantage of these definitions, is that they are scale dependent.
$\mathcal{SC}_{1}(||\mathcal{S}_{p}^{r}||)$ and $\mathcal{SC}_{2}(||\mathcal{S}_{p}^{r}||)$ depend on the radius of a sphere and $\mathcal{SC}_{1}(P_{p}^n)$ and $\mathcal{SC}_{2}(P_{p}^n)$ depend on the diffusion step of the random walker.
This scale dependence can be used to renormalize the curvature by taking the continuum limit of the discrete theory, as is done in lattice theories of quantum gravity like CDT.
In CDT the path integral over the gravitational degrees of freedom is regularized as a sum over discrete spacetimes, and a continuum limit is taken by rescaling all distances and taking the infinite-volume limit.
Our definitions are perfectly suited for these discrete spacetimes, as there is no notion of tensor calculus defined on them.

Notice that if one wants to extract the precise dimensionful value of the scalar curvature $\mathcal{R}_p$ from \eqref{eq:sc1P}, one has to know the exact linear relation between $n$ and $\tau$. To the best of our knowledge, this relation is not known in a general situation. Despite that, one can approximate this relation by the well-known, and widely used \cite{ibe2013elements}, identity obtained for a Gaussian random walk in a Euclidean space
\begin{equation}\label{eq:difftimeandstepsidentification}
    \Var\qty(S_{n}) = \sigma^2 n = 2 D \eta \tau,
\end{equation}
where $\sigma^2=Var(X_p)$ is the variance of the random variables describing each step in the random walk, $S_{n}=\sum_{i}^{n} X_i$ is the random variable given by the sum of the steps up to time $n$, $ \Var\qty(S_{n})$ is its variance, $D$ is the topological dimension of the space, and $ \eta$ is the diffusion constant of the process. Using this identification, one can obtain a numerical value for $\mathcal{R}_p$ from \eqref{eq:sc1P}.

Even though our implementations for a random walk will not be in a Euclidean space, and not use Gaussian variables, we will still use identification \eqref{eq:difftimeandstepsidentification} to determine dimensionful numerical values of  the curvature, $R_p$. We will return to this in the next sections.

\mycomment{\begin{tabular}{ |P{1.7cm}|P{16cm}|  }
\hline
\multicolumn{2}{|c|}{Examples Of Scale Dependent Observables} \\
\hline
Observable & Continuum Taylor Expansion\\
\hline
$||\mathcal{S}_{p}^{r}||$ &$(\frac{D\pi^{\frac{D}{2}}}{\Gamma(\frac{D +2}{2})})r^{D -1}(1-\frac{R_{p}}{6D}r^{2})$   \\
\hline
$(K_{p}^{\sigma})^{-1}$ &$(4\pi)^{\frac{D}{2}}\sigma^{\frac{D}{2}}(1-\frac{R_{p}}{6} \sigma)$   \\
\hline
$\frac{\bar{d}(S^{r}_{p})}{r}$ &$\small{\frac{2^{D-1}\Gamma(\frac{D}{2})^{2}}{\sqrt{\pi}\Gamma(\frac{2D-1}{2})}-( (\frac{2^{D-1}\Gamma(\frac{D}{2})^{2}}{\sqrt{\pi}\Gamma(\frac{2D-1}{2})})(\frac{D-1}{3D}) + \frac{2\pi \Gamma(\frac{1-2D}{2})\Gamma(\frac{D}{2})}{24 \sin(\frac{D\pi}{2})\Gamma(\frac{1-D}{2})\Gamma(1-D)\Gamma(\frac{D+1}{2})\Gamma(\frac{D-1}{2})}+ \frac{ \Gamma(\frac{1-2D}{2})\Gamma(\frac{-D}{2})\Gamma(\frac{D}{2})\Gamma(D+1)}{24 \sqrt{\pi}\Gamma(\frac{2-D}{2})\Gamma(\frac{D-1}{2})\Gamma(\frac{1}{2})})R_{p}r^{2}}$   \\
\hline
\end{tabular}}

\section{General scale-dependent quantities}\label{sec:gsc}
Based on the constructions presented in the previous section, in this section we generalize the two methodologies ($\mathcal{SC}_{1}$ and $\mathcal{SC}_{2}$) to extract the Ricci scalar from a more general class of scale-dependent quantities, that if computed in a Riemannian Manifold contain curvature invariant corrections. 
We will show how $\mathcal{SC}_{1}$ and $\mathcal{SC}_{2}$ still work for this more general class of quantities, even in the case were the parameters describing these quantities are random variables, and if one uses discrete notions of derivatives and integrals. 

Specifically, we consider the class of quantities $\mathcal{Q}_{p}^{s}$, that take the functional form 
\begin{equation}
	\mathcal{Q}_{p}^{s}=C_{p}s^{ \alpha_{p}}(1+\mathcal{K}_{p}(s)),
	\label{eq:12}
\end{equation}
where $p$ is a point on the space, $s$ is a distance\footnote{For example, this distance can be the geodesic radius of a sphere $r$, or the diffusion time $\tau$ of a diffusion process as we used in the previous section.} and $C_{p}, \alpha_{p} \in \mathbb{R}$.
In a Riemannian manifold a non-zero $\mathcal{K}_{p}(s)$ implies non-zero curvature of the manifold.
Examples of such quantities $\mathcal{Q}_p^s$ are: the volume of a geodesic ball or sphere, the average distance between point pairs in a geodesic ball or sphere, or the trace of the heat kernel of the covariant Laplacian.
In these cases the scale $s$ is given by the sphere/ball radius $r$, or by the diffusion time $\tau$.
For all of these quantities $\mathcal{K}_{p}(s) = 0$ in a flat manifold, but in a general Riemannian manifold this is not necessarily the case. 

The reason for considering these type of quantities is that our main goal is to apply our methods $\mathcal{SC}_{1}$ and $\mathcal{SC}_{2}$ to the quantum geometries that appear in CDT.
Since these geometries are irregular simplicial manifolds, and fractal in nature, one cannot expect that at every point $p$ volumes and return probabilities (such as $\norm{\mathcal{S}_{p}^{r}}$ and $P_{p}^n$) behave in the same manner.
This is why we generalized the quantities to be studied to have the form \eqref{eq:12}.
In the following sections of this work, we will assume that $\norm{\mathcal{S}_{p}^{r}}$ and $P_{p}^n$ can be cast within the form \eqref{eq:12}, in order to make an interpretation of the measured $\mathcal{SC}_{i}$.

The idea is to extract $\mathcal{K}_{p}(s)$ from quantities of this type, without having to worry about what $C_{p}$ and $\alpha_{p}$ are, and use our methods to define a notion of coarse-grained scalar curvature that is still valid in the geometries obtained in CDT. Notice that we accept that $C_{p}, \alpha_{p}$ and $\mathcal{K}_{p}(s)$ are point dependent. In fact, they can be randomly distributed at every point of the space.

Following the previous section, we compute each Generalized Scalar Curvature applied to $Q_p^s$
\begin{equation}\label{eq:gsc-def}
    \mathcal{SC}_{1}(Q_p^s) \coloneq  s \, \dv{s}( \dv{\log{Q_p^s}}{\log{s}} ) \,\,\,\,\,\,\,\,\,\,\,\, \mathcal{SC}_{2}(Q_p^s)
   \coloneq  \frac{1}{(\int_{0}^{s} dx \,x\,)}\int_{0}^{s} dx \,x\, \left ( \dv{\log{Q_p^s}}{\log{s}} -\dv{\log{Q_p^x}}{\log{x}} \right),
\end{equation}
keeping the point $p$ fixed. 

Since $\log{Q_p^s}=\log{C_p}+\alpha_{p}\log{s}+\log{(1+\mathcal{K}_p(s))}$, one can show that due to the distributive properties of the derivative and the integral, that hold also in an appropriate discrete implementation, 
 one can see that $\alpha_{p}$ and $C_p$ cancel out from the expressions of $\mathcal{SC}_{1}$ and $\mathcal{SC}_{2}$, and any deviation from $0$ is associated. 


Notice that it is of essence to compute \eqref{eq:gsc-def} at a fixed point $p$, otherwise the cancellation of the (possibly random) variables $C_p$ and $\alpha_p$ does not happen. In fact, if one does this calculation after an average of $Q_p^s$ at different points, one cannot guarantee that the cancellation occurs, and a bias due to the possible correlation of the variables would appear. In what follows of this work, we will compute these quantities at each point of the spaces being studied. The discrete implementations will be discussed in the following sections.

The interpretation of the presence of curvature in $\mathcal{SC}_{1}(Q_p^s)$ and $\mathcal{SC}_{2}(Q_p^s)$ applies to the quantities discussed in the previous section, $\norm{\mathcal{S}_{p}^{r}}$ and $P_{p}^n$.
In case we work in a metric measure space using the sphere volume, we use $\norm{\mathcal{S}_{p}^{r}}$ as $\mathcal{Q}_p^s$ where $r$ is used as $s$.
In the case of a metric space endowed with a random walk, we use $P_{p}^n$ as $\mathcal{Q}_p^s$ where $\tau = \frac{n \sigma^2}{2 D \eta}$ is used as $s$, assuming \eqref{eq:difftimeandstepsidentification} to hold.

Using the generalized scalar curvatures based on these quantities, we can define scalar curvature, using $\mathcal{SC}_{i}$ on a general metric space.
And we can interpret deviations from $0$ as the presence of curvature\footnote{After statistical noise and lattice artefacts are taken into account.}.

The advantage of using $\norm{\mathcal{S}_{p}^{r}}$ and $P_{p}^n$ to define coarse-grained curvatures, is that the continuum expressions in a Riemannian manifold, or at least expansions at small distance scales, are known.
So positive or negative deviations from $0$ at small distance scales can be directly related to the \emph{sign} of the scalar curvature of the manifold.

Furthermore, it is useful to compare $\mathcal{SC}_{i}$ for all length scales $s$, not just in the limit of small scales, to the results in a constantly curved Riemannian manifold.
For a $D$-dimensional constantly curved Riemannian manifold, this can be done for the quantity $\norm{\mathcal{S}_{p}^{r}}$.
For the return probability this is unfortunately impossible, because so far only asymptotic expansions are known for the heat kernel.
The full continuum expression for each $\mathcal{SC}_{i}(\norm{\mathcal{S}_{p}^{r}})$ on a $D$-dimensional sphere and a hyperbolic $D$-space can be found in appendix \ref{app:gsc-spherevol}.
Both expressions depend on 2 parameters, the dimension $D$ and the radius of the space $R$.
\mycomment{
Surprisingly, $\mathcal{GSC}_{2}(||\mathcal{S}_{p}^{r}||)$ only depends on $R$, and not on $D$.
This makes it the ideal reference to compare the results of the generalized scalar curvatures for spaces where one does not want to assume a priory a value for $D$. In some cases, one might want to keep this parameter free, determining some sort of "effective dimension" of the studied space, derived from our curvature definitions. Something similar to this was done using the QRC in 2D Euclidean dynamical triangulations in \cite{Klitgaard2018Quantizing}.}

By comparing the results of computing $\mathcal{SC}_{i}(||\mathcal{S}_{p}^{r}||)$ in a given space with the results of appendix \ref{app:gsc-spherevol}, one can determine the resemblance to a sphere or a hyperbolic space.
Furthermore, if there is a strong resemblance between the measured $\mathcal{SC}_{i}(||\mathcal{S}_{p}^{r}||)$ and the plot \ref{fig:gsc2S} for a constantly curved Riemannian manifold, an effective radius can be assigned, by fitting the measured $\mathcal{SC}_{i}(||\mathcal{S}_{p}^{r}||)$ to the formulas given in the appendix.

\section{Implementations and examples of use}\label{sec:theory-impl}
The intended use of the defined $\mathcal{SC}_{i}$ is to evaluate scalar curvature in discrete metric spaces.
In this section we define a discretization of \eqref{eq:gsc-def} and evaluate it on triangulations of two-dimensional spaces with constant curvature, the two-sphere and the two-dimensional plane.
These evaluations serve as a consistency check for the $\mathcal{SC}_{i}$, by comparing the measured curvature of the triangulations based on $\mathcal{SC}_{i}$,
with the known curvature of the smooth two-dimensional space the triangulations are approximating.
This allows us to evaluate the discretization artefacts and other limitations of our definition in a controlled setting.

\subsection{Triangulation construction}
To construct triangulations of the plane and the sphere, we will make use of so-called Delaunay triangulations of the smooth geometries.
In order to do so, we consider an embedding of the two-sphere and the plane in a three-dimensional Euclidean space, and we will restrict ourselves to discussing Delaunay triangulations is this setting.
Delaunay triangulations of a surface embedded in Euclidean space, are triangulations of a sample set of points contained in the surface, that have the property that no point in the sample is contained inside the circumcircle of any triangle of the triangulation \cite{lee1980two}.
In this work, we construct the sample set of points by using Poisson Disk sampled points \cite{Wang2020}.
Note that to create a triangulation of the plane we sample points in a square region of the plane, where we periodically tile this square regions to fill the plane, associating opposing sides.
The distance function used to create the sample of points is the induced metric on the surfaces (plane and sphere) from the embedding in euclidean space. 
Denoting the spaces being triangulated (plane and sphere) by $M$, we will denote a Delaunay triangulation of it as $\mathcal{D}_{\mathcal{T}}(M)$.
The resulting Delaunay triangulation $\mathcal{D}_{\mathcal{T}}(M)$ is a random geometry, in the sense that its vertices are randomly sampled.

For both Delaunay triangulations we construct metric spaces $(P(\mathcal{D}_{\mathcal{T}}(M)), d_{l})$ and $(F(\mathcal{D}_{\mathcal{T}}(M)), d_{d})$.
Here $P(\mathcal{D}_{\mathcal{T}}(M))$ denotes the set of points constituting the vertices of the triangulation, together with the link distance $d_{l}$ as the distance function, which is the minimal number of links between two points in the triangulation; very similar to the construction used in \cite{Klitgaard2018}.
$F(\mathcal{D}_{\mathcal{T}}(M))$ denotes the set of points constituting the triangles of the triangulation, together with the dual distance $d_d$ as the distance function, which is given by the minimal number of edges traversed from triangle to triangle.
Note that $F(\mathcal{D}_{\mathcal{T}}(M)$ can be seen as the dual of the Delaunay triangulation, a structure which is also known as a Voronoi tessalation.
Additionally, we endow $(P(\mathcal{D}_{\mathcal{T}}(M)), d_{l})$ with a Hausdorff measure $\mu_{h}$, to make it a metric measure spaces.
This Hausdorff measure is given by $\mu_{h}(A)=\#A$ for any subset $A \subseteq P(\mathcal{D}_{\mathcal{T}}(M))$, where $\#A$ is the number of elements in the set.
Also, we endow $(F(\mathcal{D}_{\mathcal{T}}(M)), d_{d})$ with a random walk $\{X_{i}\}$ for each triangle $i$ in $F(\mathcal{D}_{\mathcal{T}}(M))$.

\begin{figure}[th]
  \centering
  \subfloat[Poisson Disk Sampled Plane]{\includegraphics[width=0.45\textwidth]{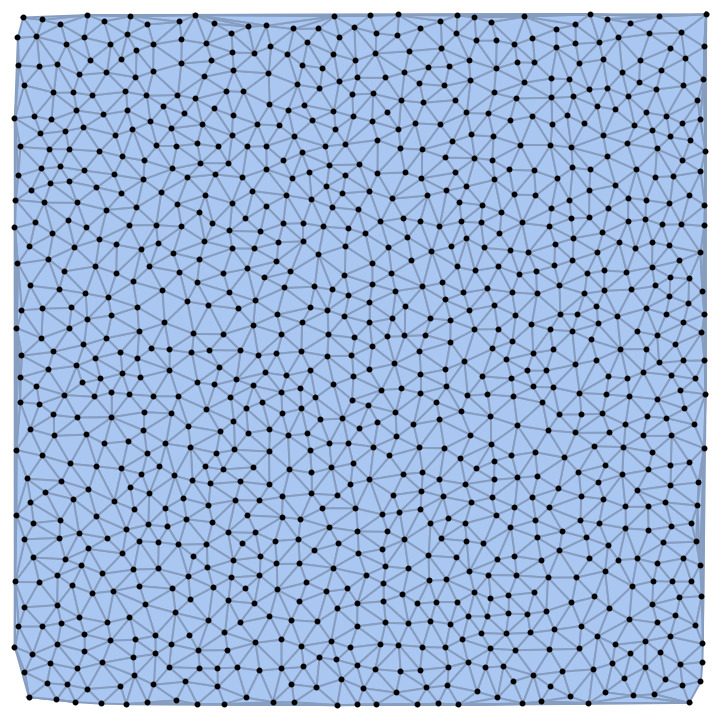}}
  \hfill
  \subfloat[Poisson Disk Sampled Sphere]{\includegraphics[width=0.45\textwidth]{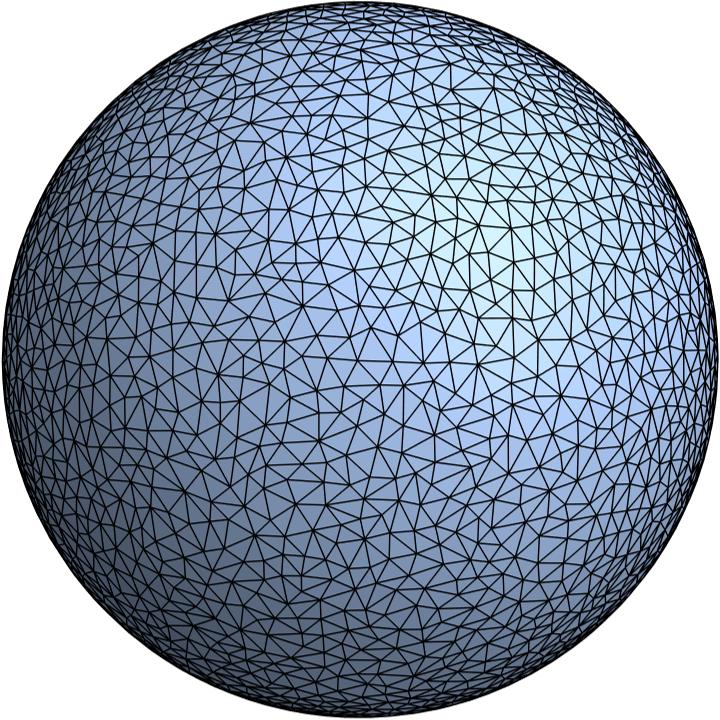}}
  \caption{Delaunay triangulations of a plane and a sphere, making use of Poisson disk sampling of points on the surface of smooth two-dimensional surfaces as described in the text. Note that we only show a Delaunay triangulation of a square region of the plane in this figure, but for the measurements we tile the square region to cover the plane.}
  \label{fig:dtriang}
\end{figure}

The reason for using the link (or dual) distance instead of the canonical distance function of the original smooth metric space, is to test the capability of our curvature definitions to recover geometrical properties of the underlying triangulated space without using the metric tensor of the underlying continuum surface.
The motivation for doing this is twofold.
First, we want to test the robustness of our curvature measures with respect to discretization artefacts.
Replacing the continuum distance function from the embedding with a lattice distance function gives a discrete approximation of the continuum distance, thus serving as a test ground for discretization artefacts.
Second, our main goal is to apply these curvature definitions to the CDT quantum geometries, where one does not even know if there is an underlying continuum manifold being approximated.
In this setting one relies solely on the discrete setting of the geometries in the path integral.
So, using the link distance in $P(\mathcal{D}_{\mathcal{T}}(M))$ also works as a preamble and test ground for implementing the generalized curvatures in those quantum geometries. 
In this controlled Delaunay triangulation setting, we expect both the link distance and dual distance to converge, up to proportionality factors, to the geodesic distance of the surfaces the triangulations are based on, in the infinite number of triangles limit.
So we expect our generalized scalar curvature definitions to recover the Ricci scalar curvature of the underlying spaces in the infinite triangle limit.
To give an idea of the type of triangulation ones obtains, we show examples of Delaunay triangulations of a plane and sphere in Fig. \ref{fig:dtriang}; note that these triangulations contain only a small number of triangles compared to the Delaunay triangulations used in our measurements.

We are interested in studying discrete spaces that are large enough to see curvature effects before reaching finite-size effects.
To this end, we studied Delaunay triangulations with close to $150\,\mathrm{k}$ vertices.
In the case of the Delaunay triangulation of the sphere, we can make a rough estimation of the effective radius in terms of the link length.
When using the link distance as distance function means setting the length of the links to be $1$, giving a triangle area of $A_\triangle = \sqrt{3} / 4$.
With a total of $N_2 = 300\,\mathrm{k}$ triangles this gives a total triangulation area (two-volume) of $A_\triangle N_2$.
Comparing this to the area of a continuum sphere $4 \pi R^2$, we obtain a rough estimate of the effective radius of $R_\text{eff} = \sqrt{A_\triangle N_2 / 4\pi} \approx 102$. 
Note that we use the Poisson disk sampling such that all links will already have very similar lengths, so the scaling to link length $1$ does not change the geometry strongly.
Still, the triangulation after rescaling the link lengths will resemble something more like a ``fuzzy'' plane or sphere, with bumps and fluctuations around the actual smooth underlying sphere. Since to use this formula, one assumes that all these fluctuations created by the ``fuzziness'' are at the same distance $R_\text{eff}$ from the origin, but giving a larger area, this will cause the estimate of the effective radius to be an overestimate of the actual radius of the underlying continuum sphere. We expect our coarse-grained notions of scalar curvature to be capable of washing this ``fuzziness'' out, and provide a more accurate (smaller) effective radius of the underlying space.

\subsection{Discrete \texorpdfstring{$\mathcal{SC}$s}{scalar curvatures}}
In order to measure the $\mathcal{SC}_{i}$ in discrete metric spaces we need to discretize the definitions \eqref{eq:gsc-def}.
For this work we ended up using the following discretizations, where now $Q_p^s$ is defined on a discrete scale $s = 0, 1, 2, \dots, s_\text{max}$, where is $s_\text{max}$ represents the last $s$ for which $Q_p^s$ is measured.
We discretize the logarithmic derivative using the simple forward finite-difference,
\begin{equation}\label{eq:logdv-discrete}
    L_p^s \coloneq \frac{\log(Q_p^{s + 1} / Q_p^s)}{\log((s + 1) / s)}.
\end{equation}
Also using the forward finite-difference for the second derivative and the cumulative sum for the integral, gives the discretized definitions
\begin{align}
    \mathcal{SC}_1(Q_p^s) &= s \qty( L_p^{s + 1} - L_p^s ), \label{eq:gsc1-def-discrete}\\
    \mathcal{SC}_2(Q_p^s) &= \frac{\sum_{k = 0}^{s} k \qty( L_p^s - L_p^k )}{\sum_{k = 0}^{s} k}. \label{eq:gsc2-def-discrete}
\end{align}
Note that \eqref{eq:logdv-discrete} is not well-defined at $s = 0$.
However, we are not interested in the scalar curvature at very small discrete length scales, because at these scales the behaviour is dominated by discretization effects, as will be seen later.
Hence, we simply leave $L_p^s$, and thus $\mathcal{SC}_1(Q_p^s)$ and $\mathcal{SC}_2(Q_p^s)$, undefined at $s = 0$, and we do not consider them in the results.
Note that this may seem to have an effect on \eqref{eq:gsc2-def-discrete} for all $s$, but this is not the case as the factor $k$ makes sure that $L_p^k$ does not contribute to the sum at $k = 0$.

In practice, we find that taking the second derivative  (more accurately, the finite difference implementation in \eqref{eq:gsc1-def-discrete}), results in the statistical fluctuations becoming very large.
So, instead of taking a difference of $L_p^s$ directly we can take the difference of some moving (weighted) average, more specifically we use
\begin{equation}\label{eq:moving-average}
    L_p^s \longmapsto \bar{L}_p^s = \sum_{k = -m}^{+m} w_k \, L_p^{s + k}, 
\end{equation}
where $m$ denotes the maximum number of points away from $s$ we take our average over, making it an average over $2m + 1$ points.
Additionally, we use weights $w_k$, which are properly normalized, $\sum_{k = -m}^m w_k = 1$.
Note that \eqref{eq:moving-average} is not defined for $s$ close to $0$ or $s_\text{max}$.
Since small scales $s$ are not interesting due to discretization artefacts, and we will explore a sufficiently large range of scales $s$, we leave the first and last points undefined and ignore them in our analysis.

\subsection{Discrete diffusion process}
Before we present the results of the discretized scalar curvatures \eqref{eq:gsc1-def-discrete} and \eqref{eq:gsc2-def-discrete} on the Delaunay triangulations of a plane and sphere, we explain the discrete implementation of the return probability $P_p^n$ measurements.
First, as mentioned before, if we want to determine precise dimensionful values for the scalar curvature, and not only its signature, we need an identification between the random walk time and the corresponding diffusion time.
 In the case of a discrete geometry, a standard implementation of a random walk is to start at a point in the space, and move to a random neighbour of that point.
The neighbour is chosen uniformly from all direct neighbours.
In the case of an irregular discrete space, a node might not have the same number of neighbours all other vertices.
Therefore, the probability distribution of the random variables determining the random walk is different for each point.

To get closer to a situation that best resembles the conditions that allow the identification \eqref{eq:difftimeandstepsidentification}, we implement the random walk in such a way that each step is equally distributed.
This is because, if the variables are not identically distributed, the variance computed in \eqref{eq:difftimeandstepsidentification} is not necessarily equal to a factor $\sigma^{2}$ times the number of steps taken.
In the case of not equally distributed random variables describing each step, relation \eqref{eq:difftimeandstepsidentification} could be regarded as an effective relation, where the proportionality factor $\sigma^{2}$ would be some effective variance of each step.
In this situation, one would have to determine this relation by other means. 

\begin{figure}
    \centering
    \includegraphics[width=0.9\textwidth]{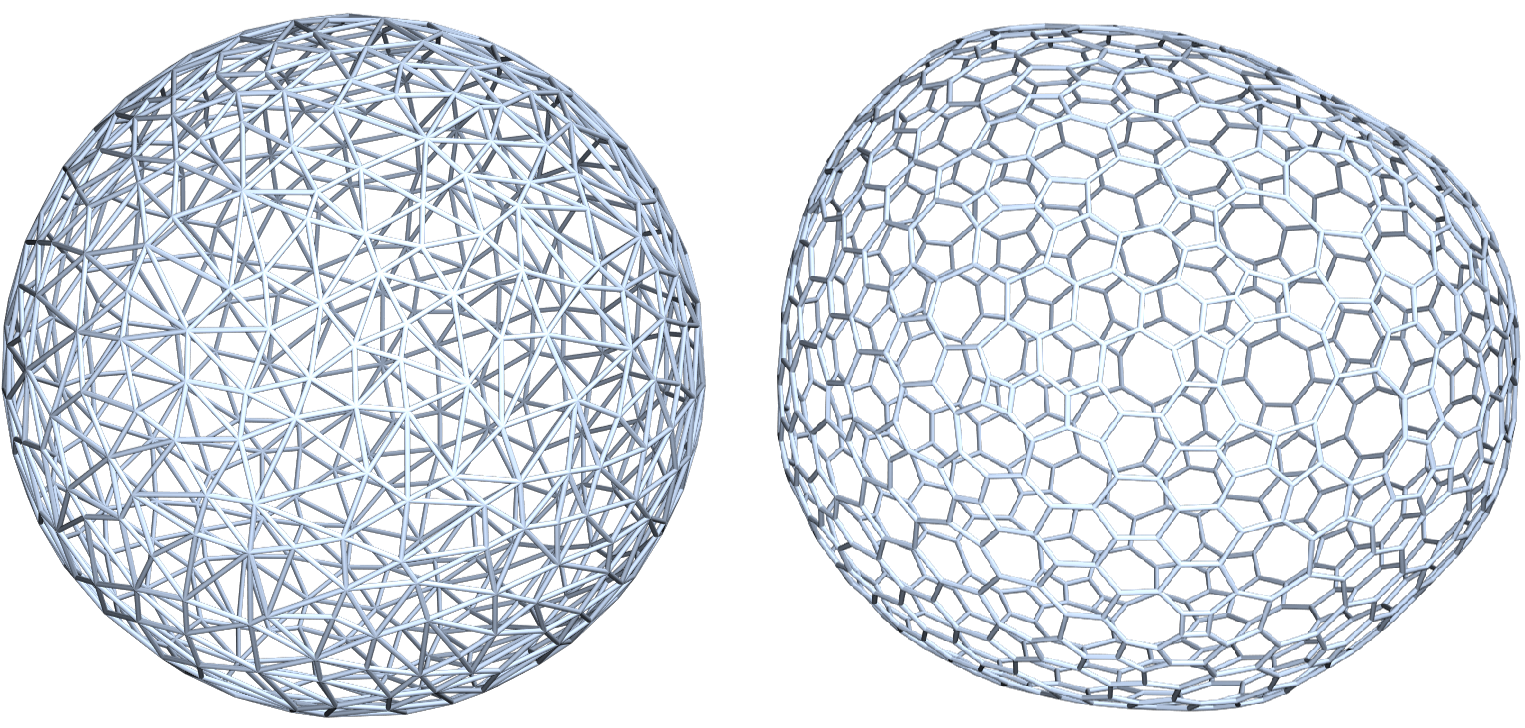}
    \caption{On the left, a Delaunay triangulation of a sphere constructed from a sample of $500$ points on a continuum sphere.
    On the right, the associated dual graph of the triangulation.
    Notice that each node in the dual graph has coordination number $3$, even though the vertex graph of the triangulation does not have a fixed coordination number.}
    \label{fig:dualeqnum}
\end{figure}

Because of this, we will restrict ourselves to a case where we can compute $\sigma^{2}$ exactly. This is the case for the metric space whose construction is based on the dual of the Delaunay triangulations, meaning the space based on the triangles with the dual distance.
In the dual graph of the triangulation, each node will have the same number of nearest neighbours, and thus, the random variables describing each step are equal to a uniform distribution among the three nearest neighbours of each node.
For a visualization of this effect on a Delaunay triangulation of a sphere and its dual graph, see Fig. \ref{fig:dualeqnum}.
If the triangulation is equilateral, ($\sigma^{2}$) can be computed exactly, as is done later.

Even though the random variables in this discrete setting are not Gaussian but uniform, we appeal to the Central Limit Theorem and assume that at some large value of $n$ the identification \eqref{eq:difftimeandstepsidentification} will hold approximately.
Therefore, we will use identification \eqref{eq:difftimeandstepsidentification} for our diffusion processes in the dual lattice, even though we are not exactly in the situation where this identification is exact.
We remark that an implementation on the vertices of the triangulation is also possible if one interested in detecting the presence of curvature using our prescriptions.
However, we wish to make use of a precise identification between diffusion steps and diffusion time, to obtain predictions for the dimensionful scalar curvature in the discrete spaces considered here.

Since in the Delaunay triangulation the edge length was set to $1$ everywhere, the edge length in the dual graph is going to be $\frac{1}{\sqrt{3}}$ so in the dual we will also have the same "fuzziness" as in the Delaunay Triangulation. From now on, when we speak of the random walk in the Delaunay triangulations, we will be speaking of a random walk on the dual graph of that Delaunay triangulation, after rescaling all the edges to length $1$. 

The discrete diffusion process we use is implemented as follows.
Let $P(p', p; n)$ be the probability of a random walker moving from point $p$ to $p'$ on the dual graph of the triangulation, in $n$ steps.
We start in our diffusion process with $P(p', p; 0) = \delta_{p, p'}$, where $\delta$ is the Kronecker delta symbol; this signifies that the walk starts at point $p$ and is normalized such that $\sum_{p'} P(p', p; n) = 1$ for all $n$;
The diffusion process is then implemented using the evolution equation \cite{Ambjorn1995FractalStructure, benedetti_spectral_2009}
\begin{equation}
    P(p', p; n + 1) = (1 - \eta) P(p', p; n) + \frac{\eta}{3} \sum_{q \sim p'} P(q, p; n),
    \label{eq:mastereqretprob}
\end{equation}
where $q \sim p'$ denotes the set of all direct neighbours of $p'$ in the graph. 
Also, $\eta \in \left(0, 1\right]$ takes the role of a discrete diffusion constant, which can be set smaller than $1$ to mitigate the large differences between the return probability for odd and even $n$ for small $n$; we use $\eta = 0.9$ in our measurements.
The return probability is determined by measuring $P_{p}^n = P(p, p; n)$ after each diffusion step.

\subsection{Discrete scalar curvature}
Using the discretizations \eqref{eq:gsc1-def-discrete} and \eqref{eq:gsc2-def-discrete}, with the discrete sphere volume $\norm{S_p^r}$ and return probability $P_p^n$ we are able to measure the scalar curvature in the Delaunay triangulation of the plane and sphere.

\subsubsection*{Sphere volume}
\begin{figure}[th]
    \centering
    \includegraphics[width=0.48\linewidth]{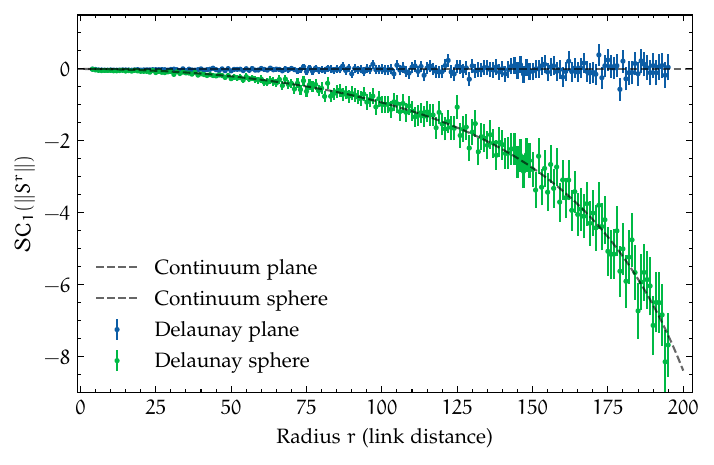}
    \hfill
    \includegraphics[width=0.48\linewidth]{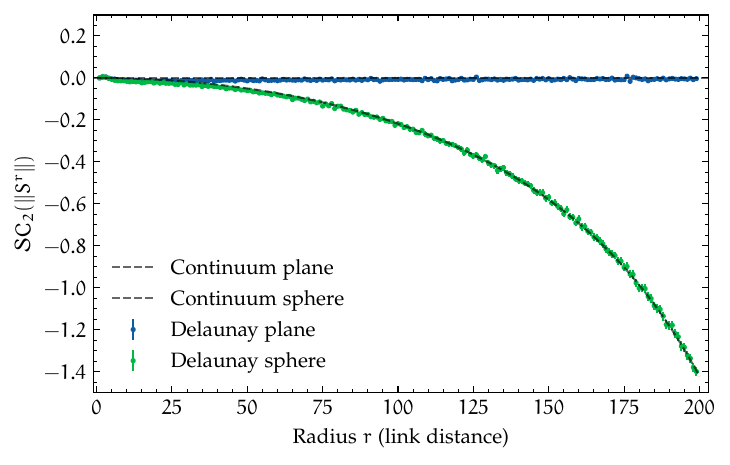}
    \caption{Scalar curvatures $\mathcal{SC}_1(\norm{S^r})$ and $\mathcal{SC}_2(\norm{S^r})$ for the sphere volume for the Delaunay sphere and plane. These results are sample averages based on $10^5$ sampled origins in each geometry.
    Included are the continuum curves of the scalar curvatures of the plane and the two-sphere, the latter of which is fitted to the Delaunay sphere results.}
    \label{fig:delaunay-svol-sc}
\end{figure}

The obtained results for $Q_p^s \to \norm{S_p^s}$ with $s \to r$, are presented in Fig. \ref{fig:delaunay-svol-sc} for both geometries.
Note that the results presented here are a sample average of $\mathcal{SC}_i(\norm{S_p^r})$ for $10^5$ uniformly sampled origins $p$ on the Delaunay geometry.
Additionally, $\mathcal{SC}_1(\norm{S_p^r})$ is computed using a moving average as described by \eqref{eq:moving-average} with $m = 5$, in this case using uniform weights $w_k = 1 / \qty(2m + 1)$.

From these results it is evident that $\mathcal{SC}_1(\norm{S^r})$ is more sensitive to discretization artefacts.
This is due to the use of the second derivative, and precisely why a moving average as given by \eqref{eq:moving-average} is used, but even with this moving average the results are significantly more noisy than for $\mathcal{SC}_2(\norm{S^r})$.
However, $\mathcal{SC}_2(\norm{S^r})$ shows lattice artefacts at small radii $r$, whereas $\mathcal{SC}_1(\norm{S^r})$ is virtually unaffected by these deviations from $0$.
The presence of lattice artefacts at larger scales $r$ is caused by the integral in the definition \eqref{eq:gsc2-def-discrete} of $\mathcal{SC}_2(\norm{S_p^r})$, which causes the lattice artefacts present in $S_p^r$ at small $r$ to have significant contributions at larger $r$.
Finally, we note that in the presented $r$ range the lowest-order, which is second-order, are clearly visible, and for large $r$ higher-order corrections are also present. This was determined by comparing to quadratic fits.
As a consequence, one has to be careful when trying to extract curvature based on the lowest-order correction, as will be discussed later.

To compare these results of the Delaunay triangulations to the continuum geometries they are based on, we compare them to the continuum expectations of the scalar curvatures.
In Fig. \ref{fig:delaunay-svol-sc} we include a fit of $\mathcal{SC}_i(\norm{S_p^r})$ in a continuum sphere, as well as the identically 0 line of a continuum plane, both displayed by a dashed line.
These are one-parameter fits as given in appendix \ref{app:gsc-spherevol} ($D = 2$), based on the radius $R$ of the two-sphere.
From this fitting procedure we obtain the estimated two-sphere radii
$R = 92.2 \pm 0.3$ and $R = 92.5 \pm 0.1$ based on $\mathcal{SC}_1(\norm{S^r})$ and $\mathcal{SC}_2(\norm{S^r})$ respectively.
It is important to remark that since the fitting functions used do not have the initial over- or undershoot present in the numerical results, the fitting itself, and therefore the estimated $R$, is more accurate at larger scales.
We note that these estimates of the radius are compatible with each other, indicating that $\mathcal{SC}_1(\norm{S^r})$ and $\mathcal{SC}_2(\norm{S^r})$ are expected to yield the same scalar curvature in the continuum limit.

The intended use of the scalar curvatures $\mathcal{SC}_i$ is to measure curvature using only (reasonably) small scales, not necessarily by fitting $\mathcal{SC}_i$ to an analytical expectation.
As such we analyse the small-scale behaviour of the measured scalar curvatures to obtain an estimation of the two-sphere radius $R$.
To define curvature using the knowledge of the continuum expansion of $\mathcal{SC}_i$, we have to be sure that we are not in a range of lattice artefacts.
For $\mathcal{SC}_1(\norm{S^r})$ and $\mathcal{SC}_2(\norm{S^r})$ we observe lattice artefacts to be negligible for $r > 25$. 
Based on the continuum formulas for $\mathcal{SC}_i(\norm{S_p^r})$ in appendix \ref{app:gsc-spherevol}, we find that for $\frac{r}{R} \leq \frac{3}{4}$, a quadratic approximation of the expansion deviates less than $3\%$ from its actual value. The expressions for each $\mathcal{SC}_i(\norm{S_p^r})$ of this relative error can be found in the appendix as well. 
So, it can be used as an approximation to compute the curvature in this range.
In this range we estimate the scalar curvature $\mathcal{R}$, as
\begin{equation}
    \mathcal{R}_p \simeq - \frac{3 D}{2} \, \frac{\mathcal{SC}_1(\norm{S_p^r})}{r^2} + \mathcal{O}(r^2)
    \qquad \text{and} \qquad
    \mathcal{R}_p \simeq - 6 D \frac{\mathcal{SC}_2(\norm{S_p^r})}{r^2} + \mathcal{O}(r^2)
    \label{eq:approxRp}
\end{equation}

We apply this approximation around $r=75$ for both $\mathcal{SC}_1(\norm{S_p^r})$, using $D=2$.
To avoid the large statistical error of a measurement at a single radius we take an estimate over the range $r \in [70, 79]$ for each $r$ and average the resulting estimates.
With the obtained estimate of the Ricci scalar $\mathcal{R}_p$, one can estimate the more intuitive effective radius $R_\text{eff}$ defined through $\mathcal{R}_p \coloneq \frac{D(D-1)}{R_\text{eff}^{2}}$, which yields an estimation for the size of the approximate sphere being triangulated.
We find the effective radii to be $R_\text{eff} = 89 \pm 9$ and $R_\text{eff} = 87 \pm 5$ based on 
$\mathcal{SC}_1(\norm{S^r})$ and $\mathcal{SC}_2(\norm{S^r})$ respectively.
The errors represent the $68\%$-confidence interval based on the propagated errors of the $\mathcal{SC}_i$ together with the systematic error of the truncation to second order, based on the results in appendix \ref{app:gsc-spherevol}.

These estimates of the sphere radius based on the first-order correction are found to be compatible with each other, and almost compatible with the estimates based on a full fitting.
However, the estimate based on $\mathcal{SC}_2(\norm{S^r})$ is found to be relatively low compared to the more accurate fitting estimates.
This may be understood by the still significant lattice artefacts around $r = 75$, despite our expectations.
It is possible that the integral in the definition of $\mathcal{SC}_2(\norm{S^r})$ causes the lattice artefact to still have a significant contribution around $r = 75$.
But at this radius the higher order corrections also have a significant contribution (around 3\%), so it seems that $\mathcal{SC}_2(\norm{S^r})$ is not ideal to estimate curvature based on a single value in this case, because of the relatively large lattice artefacts.
Finally, note that all the effective radii estimates are significantly smaller than the naive estimate of the radius of $R_\text{eff} \approx 102$ based on a total volume comparison, as was expected given that our definitions coarse grain the fluctuations induced by setting all the link lengths to $1$.
Thus, we find the effective sphere the Delaunay triangulation describes to be slightly smaller than the underlying smooth sphere, as if the triangulation is slightly crumpled up.

We find that the best of the two prescriptions is $\mathcal{SC}_{2}(\norm{S_p^r})$, because of its robustness against noise and still reasonably fast convergence towards the continuum fit of the Delaunay triangulation.
The fact that $\mathcal{SC}_{2}(\norm{S_p^r})$ only contains one derivative, makes it more robust at larger discrete scales, which is important if one desires to explore the continuum limit of a discrete geometry.
By comparing the numerical results with the continuum formulas given in appendix \ref{app:gsc-spherevol}, we find that $\mathcal{SC}_{2}(\norm{S_p^r})$ is more suitable to estimate the curvature at larger scales, and not very reliable for estimating curvature at short scales due to the presence of small but still significant lattice artefacts.

\subsubsection*{Return probability}
\begin{figure}[ht]
    \centering
    \includegraphics[width=.48\linewidth]{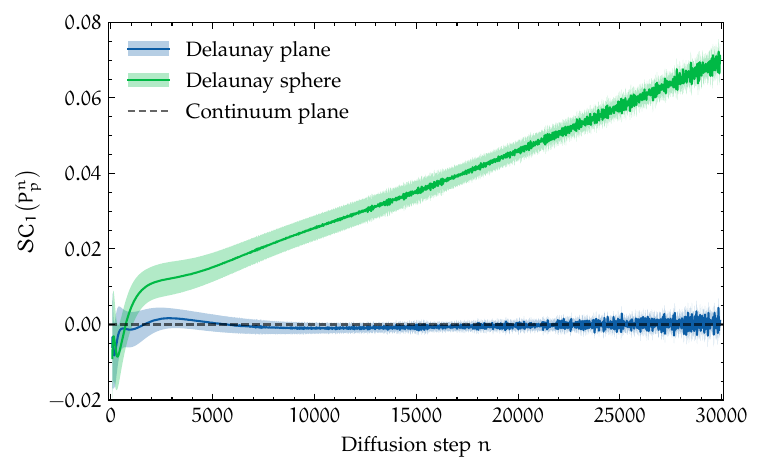}
    \includegraphics[width=.48\linewidth]{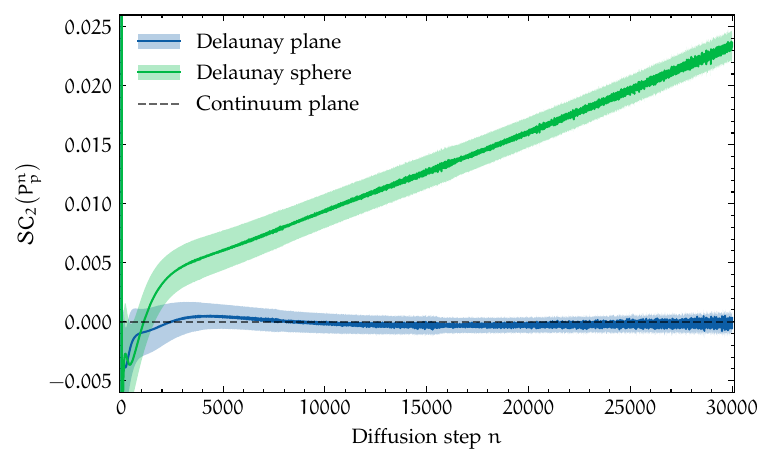}
    \caption{Scalar curvatures $\mathcal{SC}_1(P^n)$ and $\mathcal{SC}_2(P^n)$ for the return probability on the Delaunay sphere and plane. These results are sample averages based on $50$ sampled origins in each geometry.
    Included are the continuum curve of the zero scalar curvature of the plane.}
    \label{fig:delaunay-diff-sc}
\end{figure}

The obtained results for $Q_p^s \to P_p^n$ with the diffusion time identified $s \to \tau = \frac{n \sigma^2}{2 D \eta}$, are presented in Fig. \ref{fig:delaunay-diff-sc} for both geometries.
Again note that the results presented here are a sample average of $\mathcal{SC}_i(P_p^n)$ for $50$ uniformly sampled origins $p$ on the Delaunay geometry.
Additionally, $\mathcal{SC}_1(P_p^n)$ is computed using a moving average as described by \eqref{eq:moving-average} with $m = 72$, in this case using Gaussian weights $w_k = \exp(\frac{k^2}{2 (m/4)^2})$.

We studied  diffusion steps up to values $2$ orders of magnitude larger than the lattice distance used with the sphere surface.
Nonetheless, we see that in the measured range shows only the lowest-order correction to flatness, which in this case in linear.
Therefore, it seems that the diffusion process explores distance scales much smaller than those probed for the sphere volumes in Fig. \ref{fig:delaunay-diff-sc}, where higher-order corrections are already observed at the end of the plots.
Because of the already large number of diffusion steps it became computationally unfeasible to study distance scales equivalent with those of the sphere volume measurements.
However, this is no problem as both methods can be used complementary to study different scale ranges in more detail in the same metric space.
This indicates that $\mathcal{SC}_i(P_p^n)$ is ideal to define the scalar curvature at short scales, since the higher-order contributions are negligible until large diffusion times.

Given that there is no closed formula for the return probability on a sphere, only asymptotic expansions, we do not have a closed analytical expression of $\mathcal{SC}_i(P_p^n)$ to fit to.
Nevertheless, we can still estimate the curvature and the radius of the underlying sphere based on the first-order correction,
\begin{equation}
    \mathcal{R}_p \simeq \frac{12 \eta D}{\sigma^{2}} \frac{\mathcal{SC}_1((P_p^n)^{-1})}{n } + \mathcal{O}(n)
    \qquad \text{and} \qquad
    \mathcal{R}_p \simeq \frac{36 \eta D}{\sigma^{2}} \frac{\mathcal{SC}_2((P_p^n)^{-1})}{n } + \mathcal{O}(n)
    \label{eq:approxRpRetp}
\end{equation}

We apply this approximation around to the range $n \in [6000, 10000]$ for both $\mathcal{SC}_{i}(P_p^n)$, where $D = 2$, because this appears to be a range of $n$ where lattice artefacts can be neglected, and higher order correction are still negligible.
For this estimate the value of the standard deviation $\sigma^2$ in \eqref{eq:difftimeandstepsidentification} is necessary.
In the dual of a triangulation, this value can be computed exactly, to be\footnote{This can be computed as the standard deviation of a uniform distribution that can only take 3 possible values, moving at points at distance $\frac{1}{\sqrt{3}}$ from the centre.} $\sigma^{2} = \frac{1}{3}$.
We then find the effective radii to be $R_\text{eff} = 96 \pm 4$ and $R_\text{eff} = 91 \pm 4$ based on $\mathcal{SC}_1(P^n)$ and $\mathcal{SC}_2(P^n)$.
Where the errors again represent the $68\%$-confidence interval based on the propagated errors of the $\mathcal{SC}_i$.

The estimates of the effective radius made here based on the return probability turn out to be consistent with each other and with the results we obtained using the sphere volumes,
further solidifying $\mathcal{SC}_1$ and $\mathcal{SC}_2$ as a consistent method of measuring curvature.

With the return probability $P_p^n$ we think that $\mathcal{SC}_{2}(P_p^n)$ is again the best prescription to obtain the curvature of the space.
There is no significant difference with $\mathcal{SC}_{1}(P_p^n)$ regarding behaviour at large scales, but we think it is best to use $\mathcal{SC}_{2}(P_p^n)$ because it requires no moving average to obtain reasonable results, unlike $\mathcal{SC}_{1}(P_p^n)$ due to the presence of a second derivative.

Comparing our results based on the sphere volume with those based on the return probability, $\mathcal{SC}_{2}(P_p^n)$ seems to be more suitable to estimate curvature at short scales than both  $\mathcal{SC}_i(\norm{S_p^r})$.
This is because we observe the higher order corrections to be significantly smaller on a similar distance scale.
So ideally, used together, $\mathcal{SC}_{i}(P_p^n)$ will probe short-scale curvature, and $\mathcal{SC}_{i}(\norm{S_p^r})$ will probe larger distance scales, giving a larger overall range of scales at which scalar curvature can be measured in a metric space.

One last remark is in order.
We expect that all our estimations of the scalar curvature will match exactly in the limit of vanishing lattice spacing.
However, we did not perform a study of finite-size scaling of our results at different triangulation volumes, so we do not have a good understanding of how the errors will change with increasing volume at we can only conjecture they will agree in the limit.
Such a scaling analysis would be very useful in understanding the discretization effects of the presented curvature quantities, and is highly recommended being used in follow-up research.

In the following section, we will use these methods to determine the scalar curvature of the quantum geometries that appear in the path integral of 2D Causal Dynamical Triangulations.
Since these geometries are also triangulations, we use the same prescriptions as the ones used for the Delaunay triangulations in this section.

\section{Quantum scalar curvature}\label{sec:qsc}
As discussed in the introduction, the main motivation for our generalized definitions of curvature is its use in a quantum geometry setting.
Here, there is no notion of tensor calculus, but we can define distances and a diffusion process. 
In order to create valid observables for the quantum geometries they must be diffeomorphism invariant, which means that point dependence has to be eliminated.
This is because the points in the quantum geometries are fluctuating, and no identification between points in different configurations can be made.
As such, we will define an average quantum scalar curvature observable, for which we will remove the point dependence of our curvature definitions by taking a manifold average. 
To that end, we define the \emph{Quantum Scalar Curvature} $\mathcal{QSC}_i(\mathcal{Q}^s)$ based on an $s$ scale-dependent local observable $\mathcal{Q}_p^s$ to be:
\begin{equation}\label{eq:qsc-def}
    \mathcal{QSC}_i(\mathcal{Q}^s)
    = \frac{\int \dd[D]{p} \sqrt{\abs{g}} \, \mathcal{SC}_i(\mathcal{Q}_p^s)}{\int \dd[D]{p} \sqrt{\abs{g}}},
\end{equation}
where the manifold average is taken over manifold $M$ with metric $g$, where $\abs{g}$ denotes the metric determinant.
The expectation value of this quantum scalar curvature $\mathcal{QSC}_i(\mathcal{Q}^s)$ observable is given by,
\begin{equation}
    \ev{\mathcal{QSC}_i(\mathcal{Q}^s)}
    = \frac{1}{Z} \int \mathcal{D}\qty[g] \, e^{-S_\text{eu}[g]} \,
    \mathcal{QSC}_i(\mathcal{Q}^{s}),
\end{equation}
which is the Euclidean expectation value given as a Euclidean path integral over all possible geometries.
The measure $\mathcal{D}[g]$ here, denotes that the integral is taken over distinct geometries, which is given by the measure over all metrics $\mathcal{D}g$, dividing out the diffeomorphisms of the geometry \cite{ambjorn2014quantum}.
The Euclidean action $S_\text{eu}(g)$ is dictated by the particular model one is working with.
Furthermore, we denote the partition function as $Z = \int \mathcal{D}\qty[g] \, e^{-S_\text{eu}[g]}$.
For this paper we choose the setting of two-dimensional Causal Dynamical Triangulations (CDT) \cite{ambjorn20132d}.

\subsubsection*{Causal Dynamical Triangulations}
In general, CDT is a non-perturbative method to quantize gravity.
It uses lattice techniques to regularize the sum over all possible metrics on a manifold, by replacing and defining it as a sum over all possible simplicial discretizations of it, with some constraints on the possible gluings between the constituting simplices.
These constraints are the ones responsible for allowing a well-defined Wick rotation in CDT, unlike other approaches where sometimes it is not even clear if this rotation exists.
The causal dynamical triangulations in the name of the approach are the simplicial manifolds that represent curved, Lorentzian spacetimes appearing in the gravitational path integral. The causality is imposed by a distinction between time and space like edges of the simplices, and a particular way of gluing them such that at each point there is a foliation of the space in the direction of the time-like edges. By gluing these simplices, one obtains a simplicial manifold. 
Using these discrete elements one can actually make sense of the path integral over all geometries. The study of the resulting (discrete) quantum geometries is usually done by means of Monte Carlo simulations. With the use of these simulations, one can generate an ensemble of path integral configurations, where each of them will be a simplicial manifold, and therefore a discrete geometry. The expectation value of an observable over this ensemble, is later on approximated by an average over a sample of configurations in this ensemble. The ensemble of all these path integral configurations constitutes the so-called quantum geometry under study.

Since it is out of the scope of this work to properly introduce all the details of CDT, we refer the reader to some extensive reviews on the matter \cite{Ambjorn2012, Loll2019}.
What the reader should retain from this brief and incomplete introduction to CDT, is that from the CDT simulations, one can obtain an ensemble of discrete geometries constituting a quantum geometry, where the main interest is to study its geometrical properties, like curvature.

\begin{figure}[t]
    \centering
    \includegraphics[width=\linewidth]{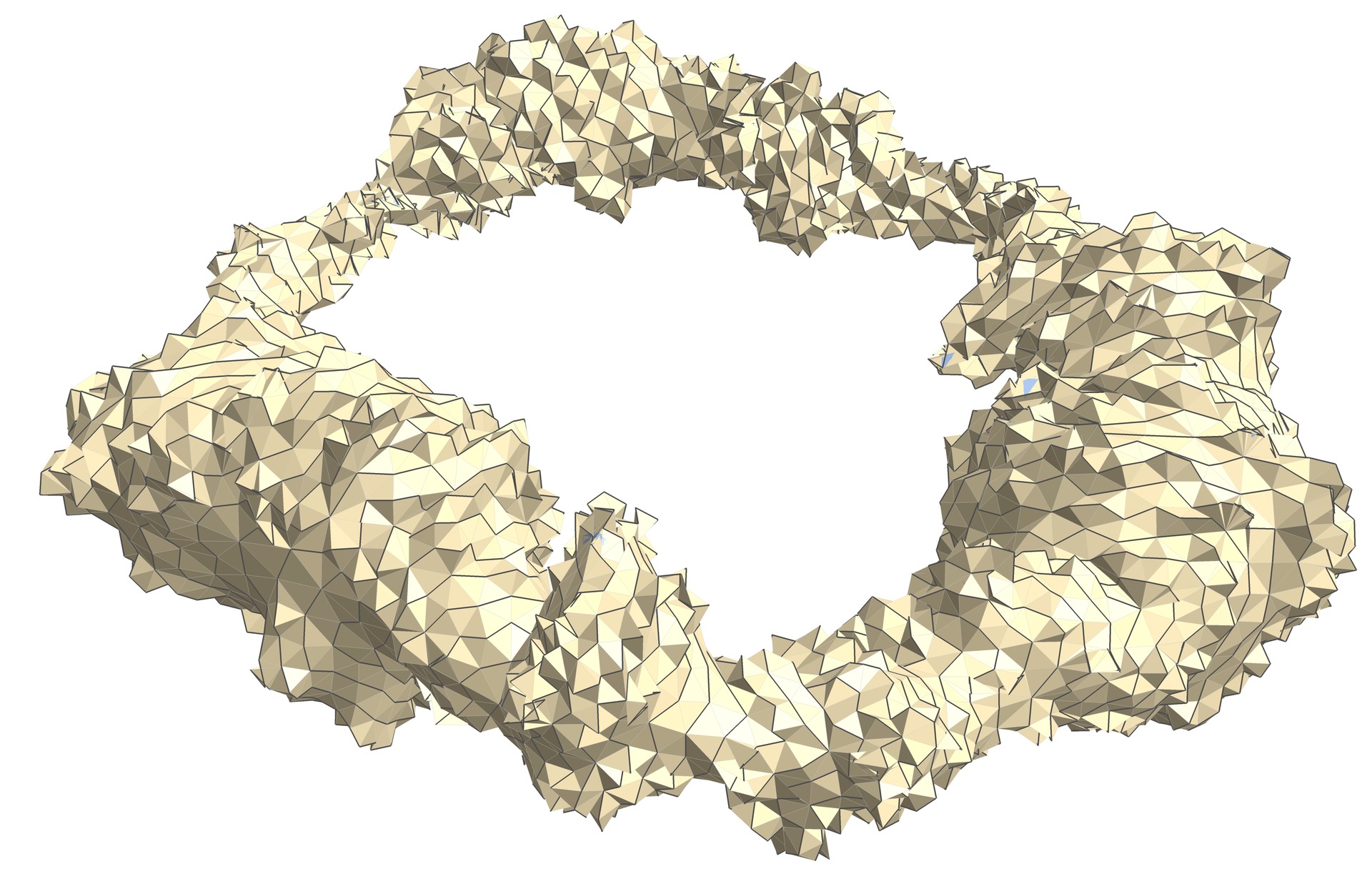}
    \caption{Visualization of a typical triangulation of 2D CDT with $N_2 = 10\,{k}$ triangles and $\tau = 123$ time-slices, based on a spring embedding. Spatial slices of constant time are marked by dark lines.}
    \label{fig:typical-cdt-triangulation}
\end{figure}
 
In two-dimensional dynamical triangulations, the discretized version of the Einstein-Hilbert action becomes very simple.
In this case the Euclidean action for a given triangulation $T$ is given by:
$S_\text{DT}(T) = \lambda N(T)$,
where $\lambda$ is a dimensionless cosmological constant and $N(T)$ is the number of triangles in the triangulation.
The partition function over the ensemble $\mathcal{T}$ is given by
\begin{equation}
    Z = \sum_{T \in \mathcal{T}} \frac{1}{C_T} e^{- \lambda N(T)},
\end{equation}
where $C_T$ is a symmetry factor suppressing highly symmetric triangulation.
In this case, the expectation value of the quantum scalar curvature is defined as
\begin{equation}\label{eq:cdt-expval}
    \ev{\mathcal{QSC}_i(\mathcal{Q}^s)}
    = \frac{1}{Z} \sum_{T \in \mathcal{T}} \frac{1}{C_T} e^{- \lambda N(T)}\,
    \frac{1}{\abs{T}} \sum_{p \in T} \mathcal{GSC}_i(\mathcal{Q}^s_p),
\end{equation}
where $\abs{T}$ is the total number of points $p$ in the vertex (or dual) graph of the triangulation.
Instead of considering the ensemble of all triangulations, we consider the ensemble of triangulations with a fixed total volume $N(T) = N_2$.
This fixed-volume ensemble is much more convenient to work with computationally, making it the standard choice for numerical CDT \cite{Ambjorn2012}. 
Furthermore, we only consider triangulations with a toroidal topology without boundaries.
A typical CDT configuration of this toroidal topology looks like Fig. \ref{fig:typical-cdt-triangulation}.

As mentioned before, we estimate the quantum scalar curvature by sampling triangulations from the ensemble using Markov chain Monte Carlo methods.
It is possible to compute the triangulation average of $\mathcal{GSC}_i(\mathcal{Q}^s_p)$ in each triangulation, but computationally rather expensive.
Instead, we estimate the triangulation average with a sample average, based on a uniform sample of points.
To decrease the effect caused by the compactness of the geometry we perform the actual measurements on an unwrapped version of the simulation toroidal triangulations.
Specifically we cut the triangulations open along two closed non-contractible loops and tile the unfolded triangulation to create an infinite periodic triangulation of planar topology; for more details see \cite{DuinMasterThesis}.

Identical to the implementation in the Delaunay triangulations, the sphere volume measurements are implemented by counting the number of points in vertex graph at a given link distance $r$ from the origin.
The computation of the return probability of a random walker is also implemented identically to the Delaunay triangulations.
We use the dual graph of the CDT triangulations and perform the random walk by updating the probability of the random walk with \eqref{eq:mastereqretprob}.
The return probability at a point $p$ after $n$ steps in a configuration $T$, is computed by recording the value of $P_T(p, p; n)$ after each diffusion step in \eqref{eq:mastereqretprob}.

\subsubsection*{\texorpdfstring{$\mathcal{QSC}$}{QSC} results}
\begin{figure}[th]
    \centering
    \includegraphics[width=0.495\linewidth]{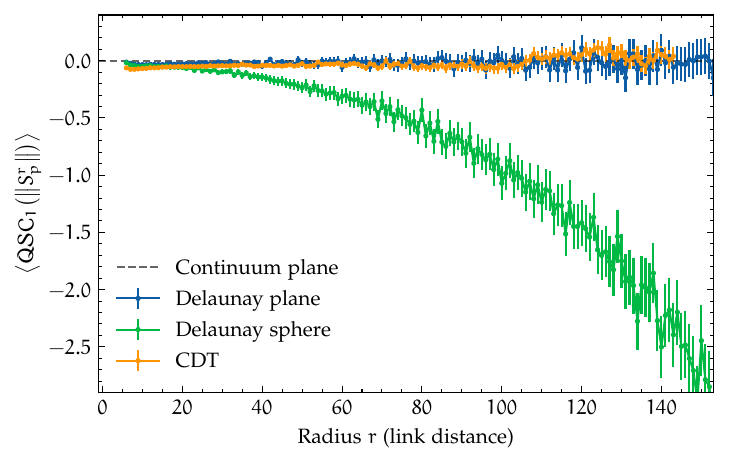}
    \includegraphics[width=0.495\linewidth]{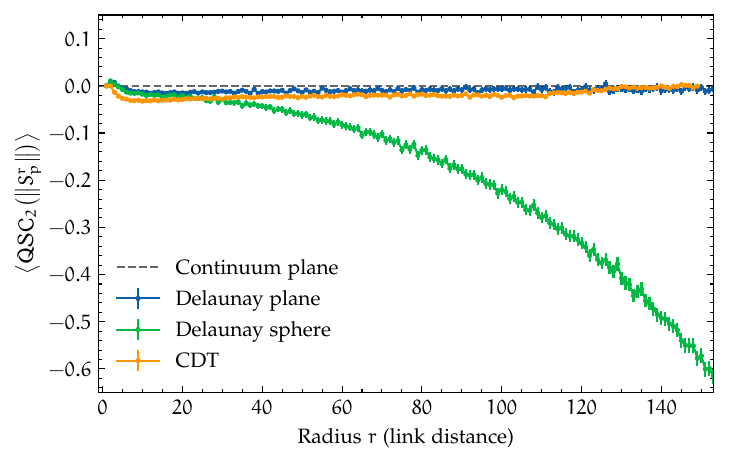}
    \caption{The quantum scalar curvatures of the sphere volume  $\ev{\mathcal{QSC}_i(\norm{S_p^r})}$ on 2D CDT with $N_2 = 300\,\text{k}$ triangles and $\tau = 219$ time-slices.
    As a comparison the generalized scalar curvature results $\mathcal{SC}_i(\norm{S_p^r})$ of the Delaunay triangulations as already presented in Fig. \ref{fig:delaunay-svol-sc} are also included with the same number of triangles, $300\,\text{k}$.
    The error bars represent $95\%$-confidence intervals; for CDT this is based on the fluctuations in $\mathcal{QSC}_i$ over $5000$ CDT configurations.}
    \label{fig:qsc-cdt-spherevol}
\end{figure}

Using the described methods we measured the quantum scalar curvatures for sphere volumes $\norm{S_p^r}$ in CDT.
We performed measurements on $5000$ triangulations to approximate the ensemble average and approximated the manifold average by a sample average over $500$ points.
The resulting quantum scalar curvatures are displayed in Fig. \ref{fig:qsc-cdt-spherevol}.
Note that just like was done for the Delaunay triangulations, $\mathcal{QSC}_1(\norm{S^r})$ is computed using a moving average as described by \eqref{eq:moving-average} with $m = 5$, using uniform weights $w_k = 1 / \qty(2m + 1)$.

From Fig. \ref{fig:qsc-cdt-spherevol} we see some interesting features of the resulting quantum scalar curvatures.
Both $\ev{\mathcal{QSC}_1(\norm{S_p^r})}$ and $\ev{\mathcal{QSC}_2(\norm{S_p^r})}$ exhibit a plateau, meaning that they are compatible with a flat curvature.
We also observe that the deviation from $0$ at small $r$ is much more significant than for the Delaunay triangulation, indicating stronger lattice artefacts.
This is to be expected as the geometries of 2D CDT exhibit a lattice structure that is significantly less representative of a smooth geometry than the Delaunay triangulations.
One may also notice that $\ev{\mathcal{QSC}_2(\norm{S_p^r})}$ is smaller than $0$ for most of the measured range of $r$, and it seems to show a slow increase toward $0$ for larger $r$.
First, note that radii $r > 100$ will be subject to finite-size effects, that is to say that these radii are large enough to cover a two-volume of the order of the total volume of the geometry and can thus have behaviour that need not be reflective of the continuum $r \rightarrow 0$ behaviour of interest.
Thus, we should not consider this region to be of interest when considering `quasi-local' coarse-grained curvatures.
Also, this deviation from $0$ for CDT is significantly smaller than that of the Delaunay triangulation of a constant curvature sphere with the same number of triangles.
For short scales ($r<25$) we expect, just as with the Delaunay triangulations, that the discretizations of derivatives and especially integrals, have significant deviations from the continuum implementations, so we don't take into account those points. 
In conclusion, both $\mathcal{QSC}_i(\norm{S^r})$ indicate that, at least within the measured scales, the ground state of the 2D CDT path integral exhibits the properties of a space with zero average curvature, when extracting curvature using the geodesic sphere surface.

We want to stress that these results are not trivial and are not to be expected a priori. The non-triviality of curvature observables was measured and noticed, for example, in 2D CDT in \cite{Brunekreef2021}.
There could be several reasons for this.
The generalized scalar curvatures we are considering here are coarse-grained quantities in the sense that they depend on the triangulation in the neighbourhood of the point $p$, with a size associated with the scale $s$ (in this case $r$).
Because of this coarse-graining, the obtained results are not necessarily the same as for the Ricci scalar, which is a truly local quantity \cite{loll2023curv}.
Even if the average of the Ricci scalar is zero on the whole manifold, since all our ensemble configurations have toroidal topology, higher order scalar curvature invariants might not.
The surprising finding in our work is that at least up to the measured scales, all of them seem to be zero.
In general, the properties of such a coarse-grained curvature is non-trivial.
These properties cannot be directly derived from the topological properties of the ensemble constituents, and can thus exhibit a non-trivial curvature average.
If one is measuring a local quantity like the Regge curvature, the result of its average over the manifold is trivial in two dimensions, because it is fixed by the topology by the Gauss-Bonnet theorem \cite{Ambjorn2012}, and is 0 for the toroidal topology we use here.
Nevertheless, higher order curvature invariants contribute to our coarse-grained observables, and non-trivial behaviour is to be expected.

\begin{figure}[th]
    \centering
    \includegraphics[width=0.495\linewidth]{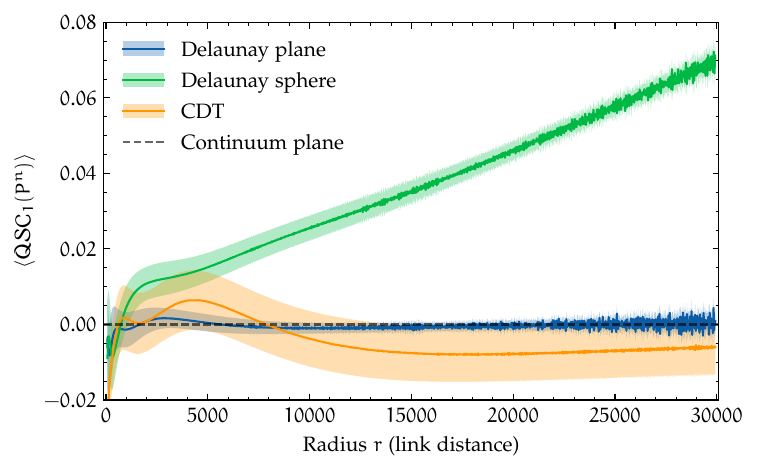}
    \includegraphics[width=0.495\linewidth]{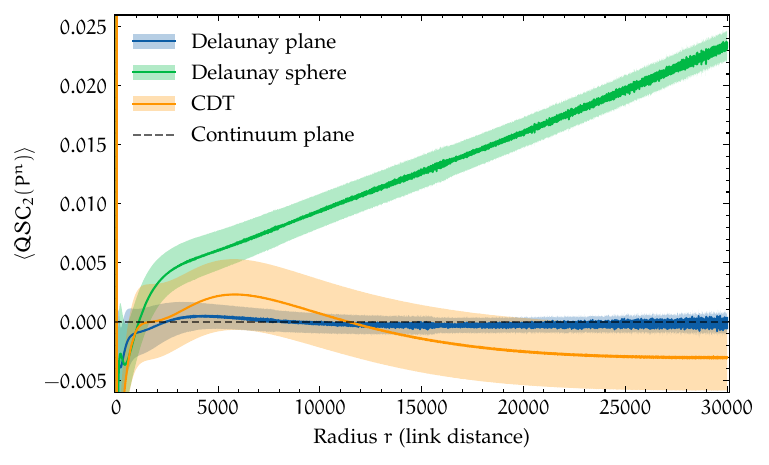}
    \caption{The quantum scalar curvatures of the return probability $\ev{\mathcal{QSC}_i(P^n)}$ on 2D CDT with $N_2 = 300\,\text{k}$ triangles and $\tau = 219$ time-slices.
    As a comparison, the generalized scalar curvature results $\mathcal{SC}_i(\norm{S_p^r})$ of the Delaunay triangulations as already presented in Fig. \ref{fig:delaunay-diff-sc} are also included with the same number of triangles, $300\,\text{k}$, as well as the continuum result for generalized scalar curvature in a plane.
    The error bands represent $95\%$-confidence intervals; for CDT this is based on the fluctuations in $\mathcal{QSC}_i$ over $1370$ CDT configurations, and for the Delaunay triangulations this is based on the fluctuation between $50$ different points on the triangulation.}
    \label{fig:qsc-cdt-diffusion}
\end{figure}

Similarly, we performed measurements of the quantum scalar curvature based on the return probability $P^n_p$ is CDT, measured using the previously described methods.
The ensemble average of \eqref{eq:cdt-expval} is estimated using $1370$ triangulations with $N_2 = 300\,\text{k}$ triangles and $\tau = 219$ time-slices.
The manifold average $\mathcal{QSC}_i(P^n)$ is approximated by a single sample of $\mathcal{SC}_i(P^n_p)$ at a point $p$ in each triangulation.
The resulting $\mathcal{QSC}_i(P^n_p)$ are presented in Fig. \ref{fig:qsc-cdt-diffusion}.
Again identically to the Delaunay triangulations $\mathcal{QSC}_1(\norm{P^n})$ is computed using a moving average as described by \eqref{eq:moving-average} with $m = 72$, using Gaussian weights $w_k = \exp(\frac{k^2}{2 (m/4)^2})$.

Similar to the sphere volume measurements we see that after a region of lattice artefacts both $\ev{\mathcal{QSC}_i(P^n)}$ are compatible with $0$ average curvature within statistical error, again indicating a flat geometry.
Note that the errors appear to be very large compared to the smoothness of the curve.
This is the case because the scalar curvature $\mathcal{SC}_i(\norm{P_p^n})$ at a single point $p$ is already relatively smooth, however the $\mathcal{QSC}_i(\norm{P^n})$ varies strongly from configuration to configuration in the CDT path integral, yielding a smooth average with large errors. 
Nevertheless, we also observe that there are large overshoots for both $\mathcal{GSC}_i(P^n)$ at $n < 1000$.
In particular, the deviations are of the size of that of the Delaunay triangulation of a sphere with the same number of triangles, meaning that the lattice artefacts of this order are large.
Despite this, at $n > 5000$ these effects seem to be washed out, and results compatible with flatness were found.
However, we acknowledge that the statistical error is large, so our results are also compatible with a non-zero curvature of CDT, albeit significantly smaller than that of the Delaunay triangulation with the same number of triangles.

Summarizing, we conclude that all our definitions of scalar curvature indicate that the path integral ground state of 2D CDT has average curvature equal to $0$.
This is different from the findings in \cite{Brunekreef2021}, where the coarsed-grained notion of curvature called Quantum Ricci curvature (QRC) is used.
Here the QRC measured in 2D CDT seemed to be non-constant, which they acknowledge to be difficult to interpret, but may seem to indicate negative curvature.
Although they also note that this curvature may vanish for scales larger scales than the ones studied in \cite{Brunekreef2021}. 
Despite the remarkable precision in their measurements, the QRC still has the problem of the unknown leading prefactor multiplying QRC, as mentioned in the introduction.
The problem in interpreting the results of the QRC, is that even if a non-constant region of the scale-dependent QRC is observed that seems to flatten out, as it appeared to do in \cite{Brunekreef2021}, it cannot be known how far away from the asymptote one is. 
Therefore, the actual numerical value of the QRC is meaningless unless one knows the multiplying prefactor, and because of that, one cannot know if the results are actually converging to it or diverging from it.
We believe to have solved this issue with our generalized curvature scalars, by eliminating this unknown leading constants and powers.
This allows us to measure scalar curvature as a deviation from $0$, a known reference value. 
This makes the interpretation of the results easier than in the presence of unknown leading powers or multiplying constants, since any deviation from $0$ can be directly interpreted as the presence of curvature.

\section{Conclusions}\label{sec:conclusion}
In this paper we define and implement generalized notions of scalar curvature, for metric spaces endowed with a measure and/or a random walk,
with the aim of providing computationally feasible methods of determining curvature that can be implemented in the non-perturbative approach to quantum gravity called Causal Dynamical Triangulations (CDT). The definitions are successfully implemented in 2D CDT. 
This is done without the need of defining or computing intermediate, more complex, tensorial quantities like the Riemann or Ricci tensor.
We do not assume the existence of tensor calculus at any stage in our definitions of curvature in these spaces.
Moreover, our constructions do not assume that the considered spaces are smooth in any sense.
Indeed, they are designed to be used in discrete and fractal spaces, like the ones appearing in CDT.
Furthermore, our definitions are intentionally coordinate free by construction, since this is situation that is encountered in the geometries of CDT. 

For the construction of these generalized notions of scalar curvature $\mathcal{SC}$, we use quantities that can be computed both in a Riemannian manifold, and in a metric space with a measure and/or a random walk.
In particular, we use the volume of a sphere $\norm{S_p^r}$ with a radius $r$, centred at a point $p$, and we use the return probability $P_p^n$ of a random walker from point $p$ after $n$ steps.
In a Riemannian manifold, both of these quantities define the Ricci scalar curvature by the leading-order scale-dependent correction from their calculation in Euclidean space.
Summarizing, the procedure used to define some $\mathcal{SC}$ is the following: 1) Find a scale-dependent quantity that can be computed both in a Riemannian manifold and in a metric space with a measure and/or a random walk, and it should be the class given the functional form \eqref{eq:12}.
The quantities we consider here, the sphere volume $\norm{S_p^r}$ and the return probability $P_p^n$, are of this form.
2) Apply a combination of integrals and derivatives to this scale-dependent quantity, to extract the leading order curvature correction in a Riemannian manifold.
3) Replicate these calculations on the metric space in question, and use the result as a definition of scalar curvature, associated with this quantity. 

We introduce $2$ different methods $\mathcal{SC}_1$ and $\mathcal{SC}_2$ given in \eqref{eq:gsc-def}, that involve different combinations of integrals and derivatives.
Their expansions in a Riemannian manifold, using $\norm{S_p^r}$ and $P_p^n$, are given in \eqref{eq:expscSphere} and \eqref{eq:expscRetP} respectively.
The reason for introducing two methods is that, in a discrete setting, different methods introduce different undesired numerical effects like noise and lattice artefacts.
The methods are compared in order to find the one that has the most desirable features, which can be different in a different setting.
We tested these definitions on random geometries generated by constructing Delaunay triangulations of a sphere and a periodic plane,
and we find that applying these prescriptions to determine the generalized scalar curvature of the corresponding Delaunay triangulations, gives consistent results with the Ricci scalar of the underlying Riemannian manifolds.
Hence, we conclude that this is a confirmation that our generalized notions of curvature can indeed be applied reliably to discrete metric spaces, and that their result can be interpreted as a notion of scalar curvature.

To determine the curvature of the ground state of the path integral in 2D Causal Dynamical Triangulations, we also measured the generalized notions of scalar curvature in this setting.
In the quantum gravity setting we extend our definitions to a diffeomorphism invariant quantum observable $\mathcal{QSC}$, by averaging over all points in the geometry \eqref{eq:qsc-def}.
We estimated the expectation value of the $\mathcal{QSC}_i$ associated with $\norm{S_p^r}$ and $P_p^n$, and determined that the expectation value of all curvature definitions are compatible with flatness; the results are given Fig. \ref{fig:qsc-cdt-spherevol} and \ref{fig:qsc-cdt-diffusion}.
This seems to suggest that the ground state of the 2D CDT path integral is flat.
We cannot make a clear physical interpretation of this result because 2D CDT has no classical limit.
In fact, the system is quantum at all scales, and to the best of our knowledge, there is no analogue in our universe that can be modelled with the quantum geometries obtained in this setting.
Since 2D CDT is a toy model for the more complex quantum geometries that emerge in 4D CDT, we consider the successful application of our methods in 2D as a first step to implement them in the physically interesting setting of 4D CDT.

We emphasize that all our generalized scalar curvatures are equivalent in the sense that they all define the Ricci scalar curvature at small scales.
There is no fundamental difference between them, only technical differences.
In the discrete cases we considered the different methods do show different discretization effects, and we find that one of our definitions of curvature is better than the other.
Based of our findings on the Delaunay triangulation we conclude that the best definition of curvature is $\mathcal{SC}_{2}(\norm{S_p^r})$ for larger distance scales.
Additionally, we conclude that $\mathcal{SC}_{2}(P_p^n)$ also provides a good prescription to obtain the curvature of the space on shorter scales, as the higher-order corrections are significantly smaller.
Given that our definitions are scale-dependent, in the sense of defining a notion of curvature at a given distance scale on the space, they can be renormalized.
This is extremely useful when the continuum limit of a space is desired, as is the situation of CDT.
Here one studies the scaling of the results when changing the size of the universe itself, after performing calculations on a fixed-size discrete triangulations
This was not done in our calculations since we were virtually using an infinite size 2D universe, by using a periodically identified section of the plane.
Therefore, our results are already approximately equal to the ones in an infinite size 2D CDT universe, and there was no need of rescaling.
Nevertheless, it would be interesting to study the dependence of our results with the size of the system in question.
As a matter of fact, even in the case of a Delaunay triangulation of a 2-sphere, one could study the convergence of the estimated scalar curvature using the different methods proposed here.
We expect that at least in a classical space like the Delaunay triangulations, the numerical differences are just a discretization artefact, and once the limit of vanishing lattice spacing is taken, all prescriptions converge to the same Ricci scalar curvature value.
But, we acknowledge that this convergence should in principle be studied to verify this claim, and a study of convergence is strongly recommended in a follow-up study.

Aside from our particular implementations and test spaces, the defined $\mathcal{SC}_i$ are ideal for the calculation of $n$-point functions.
The reason is that they explicitly define scalar curvature, without additive or multiplicative unknown constants.
Furthermore, our definitions allow us to determine more than just the presence or the absence of curvature on a space.
Since there are no unknown multiplicative constants in our observables, we can extract the scalar curvature of the given space, under some approximations explained in the previous sections.
This was done for the Delaunay triangulation of the sphere, obtaining results compatible with the underlying space being approximated by the triangulations
In future research, it would be interesting to apply these definitions to determine $n$-point functions of the scalar curvature in CDT, obtaining a non-perturbative result for the correlation function of the scalar curvature in quantum gravity.

\section*{Acknowledgements}
The authors would like to thank Renate Loll for many useful discussions and for her detailed feedback during several iterations of this manuscript. 

\appendix
\section{\texorpdfstring{$\mathcal{SC}_{i}(\norm{\mathcal{S}_{p}^{r}})$}{GSC(S)} in constant curvature manifolds}\label{app:gsc-spherevol}
In the case where the underlying space is a D-sphere of radius $R$, one can compute exactly the functional form of $\mathcal{SC}_{i}(||\mathcal{S}_{p}^{r}||)$ as a function of $r$, $R$ and $D$. This case is interesting since it could be used for direct comparison of $\mathcal{SC}_{i}(||\mathcal{S}_{p}^{r}||)$ of a general metric space with that of a D-sphere.

On a D-sphere, one can compute the surface $||\mathcal{S}_{p}^{r}||$ of a geodesic sphere of radius $r$ on top of it
\begin{equation}
	||\mathcal{S}_{p}^{r}||= 2 \pi A_{D}  R^{D-1} \sin^{D-1} \left(\frac{r}{R}\right),
\end{equation}
where $A_{D}$ is a dimensional dependent constant of the form
\begin{equation}
	A_{D}=\prod_{i=2}^{D-1}(\int_0^{\pi} \sin^{D-i}(x)dx).
\end{equation}

Using this, one can compute exactly all the generalized curvature definitions
\mycomment{\begin{equation}
	\mathcal{GSC}_{\mathcal{D}}(||\mathcal{S}_{p}^{r}||)=1 + (D-1)\,(\frac{r}{R})\,\cot\left(\frac{r}{R}\right),
\end{equation}
\begin{equation}
	\mathcal{GSC}_{\mathcal{I}}(||\mathcal{S}_{p}^{r}||)=\frac{D\,\frac{r}{R}\,\cot\left(\frac{r}{R}\right)}{\sqrt{\cos^2\left(\frac{r}{R}\right)}\,{}_2F_1\left(\frac{1}{2},\frac{D}{2},\frac{2+D}{2},\sin^2\left(\frac{r}{R}\right)\right)},
\end{equation}}
\begin{equation}
	\mathcal{SC}_{1}(||\mathcal{S}_{p}^{r}||)=(D-1)\,(\frac{r}{R})\,\left(\cot(\frac{r}{R}) - (\frac{r}{R})\,\csc^2(\frac{r}{R})\right),
\end{equation}
\begin{equation}
\begin{aligned}
\mathcal{SC}_{2}(||\mathcal{S}_{p}^{r}||)=(D - 1) & \left( \frac{{\left(\frac{r}{R}\right) \left(3\cot\left(\frac{r}{R}\right) - 2i\right) - 6\log\left(1 - e^{-2i\left(\frac{r}{R}\right)}\right)}}{3} \right. \\
& \left. - \frac{{2i \operatorname{Li}_2\left(e^{-2i\left(\frac{r}{R}\right)}\right)}}{\left(\frac{r}{R}\right)}  +  \frac{{\zeta(3) - \operatorname{Li}_3\left(e^{-2i\left(\frac{r}{R}\right)}\right)}}{\left(\frac{r}{R}\right)^2} \right),
\end{aligned}
\label{aeq:gsc2s}
\end{equation}
where $\operatorname{Li}_n$ is the Polylogarithm. 

Notice that the dimensional dependence of each of the $\mathcal{SC}_{i}(||\mathcal{S}_{p}^{r}||)$ appears to be just a multiplicative factor. This allows for a simple comparison with fractal like geometries, like the ones appearing in 2D euclidean dynamical triangulations, where the ground states shares some properties with a 5-dimensional sphere.

Now, consider a hyperbolic $D$-space with scalar curvature equal to $\frac{D(1-D)}{R^{2}}$, where we define the radius $R$ to be such that this holds.
In this case, one can also compute the surface of a sphere $||\mathcal{S}_{p}^{r}||$, where one obtains 

\begin{equation}
	||\mathcal{S}_{p}^{r}||= 2 \pi A_{D}  R^{D-1} \sinh^{D-1} \left(\frac{r}{R}\right),
\end{equation}
where it differs from the sphere case since there is a hyperbolic sine function instead of a sine function. 

\begin{figure}[t]
    \centering
    \includegraphics[width=\textwidth]{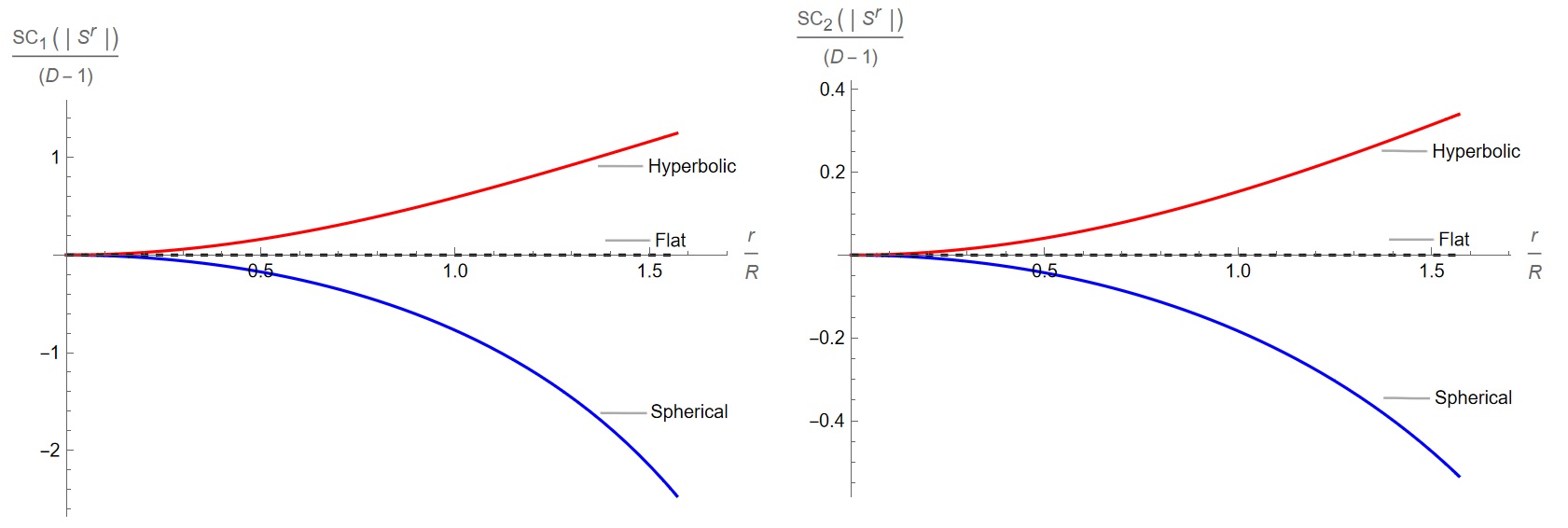}
    \caption{$\mathcal{SC}_{1}(||\mathcal{S}_{p}^{r}||)$ and $\mathcal{SC}_{2}(||\mathcal{S}_{p}^{r}||)$ for a D-Sphere of scalar curvature  $\frac{D(D-1)}{R^{2}}$, a hyperbolic D-space of scalar curvature  $\frac{D(1-D)}{R^{2}}$, and euclidean D-dimensional space.}
    \label{fig:gsc2S}
\end{figure}

Using this, one can compute exactly all the generalized curvature definitions
\mycomment{\begin{equation}
	\mathcal{GSC}_{\mathcal{D}}(||\mathcal{S}_{p}^{r}||)=1 + (D-1)\,(\frac{r}{R})\,\coth\left(\frac{r}{R}\right),
\end{equation}
\begin{equation}
	\mathcal{GSC}_{\mathcal{I}}(||\mathcal{S}_{p}^{r}||)=\frac{D\,\frac{r}{R}\,\cot\left(\frac{r}{R}\right)}{\sinh\left(\frac{r}{R}\right)\,{}_2F_1\left(\frac{1}{2},\frac{D}{2},\frac{2+D}{2},-\sinh^2\left(\frac{r}{R}\right)\right)},
\end{equation}}
\begin{equation}
	\mathcal{SC}_{1}(||\mathcal{S}_{p}^{r}||)= (D-1)\,(\frac{r}{R})\,(\coth\left(\frac{r}{R}\right) - (\frac{r}{R})\,\csch^2\left(\frac{r}{R}\right)),
\end{equation}
\begin{equation}
\begin{aligned}
\mathcal{SC}_{2}(||\mathcal{S}_{p}^{r}||)=(D - 1) & \left(  (\frac{r}{R})(\frac{2}{3} + \coth(\frac{r}{R})) - 2\log\left(e^{2\left(\frac{r}{R}\right)} - 1\right)-2i\pi \right. \\
& - \frac{2\operatorname{Li}_2\left(e^{2\left(\frac{r}{R}\right)}\right)}{\left(\frac{r}{R}\right)}  + \left. \frac{\operatorname{Li}_3\left(e^{2\left(\frac{r}{R}\right)}\right) - \zeta(3)}{\left(\frac{r}{R}\right)^2} \right)
\end{aligned}
 \label{aeq:gsc2h}
\end{equation}

\mycomment{\begin{equation}
	\mathcal{GSC}_{3}(||\mathcal{S}_{p}^{r}||)=\frac{\sqrt{\cosh^2\left(\frac{r}{R}\right)}\,{}_2F_1\left(\frac{1}{2},\frac{D}{2},\frac{2+D}{2},-\sinh^2\left(\frac{r}{R}\right)\right)\left((D-1) + (\frac{r}{R})^{-1}\,\tanh\left(\frac{r}{R}\right)\right)}{D},
\end{equation}
\begin{multline}
	\mathcal{GSC}_{4}(||\mathcal{S}_{p}^{r}||)=\frac{1}{D  \sinh^{D+1}\left(\frac{r}{R}\right) }\Big{[} \left((\int_0^{(\frac{r}{R})}\frac{\sinh^{D-1}\left(y\right)}{y}\,dy) \sinh^2\left(\frac{r}{R}\right) + \sinh^{D+1}\left(\frac{r}{R}\right)\right) \\
	\times \left(D  - (\frac{r}{R})^{-1}\,\sqrt{\cosh^2\left(\frac{r}{R}\right)}\,{}_2F_1\left(\frac{1}{2},\frac{D}{2},\frac{2+D}{2},-\sinh^2\left(\frac{r}{R}\right)\right)\tanh\left(\frac{r}{R}\right)\right)\Big{]} ,
\end{multline}
\begin{multline}
	\mathcal{GSC}_{5}(||\mathcal{S}_{p}^{r}||)=1 +\frac{(D-1)^2}{12 }  (\frac{r}{R}) \csch^2\left(\frac{r}{R}\right) \Big{[}6 (\frac{r}{R})^2 + \pi^2 + 6 (\frac{r}{R}) (1 - 2 i \pi )  + 6 (\frac{r}{R}) \cosh\left(\frac{2r}{R}\right)\\
      - 12 (\frac{r}{R}) \log(-1 + e^{\frac{2r}{R}}) - 6  \text{PolyLog}_2\left(e^{2(\frac{r}{R})}\right)\Big{]},
\end{multline}
where $\int_0^\frac{r}{R}\frac{\sinh^{D-1}\left(y\right)}{y}\,dy$ was kept in integral form, but can be computed exactly for any $D \geq 2$.}

Again, one can see that the dimension only enters as a multiplicative factor.  For the sake of completeness, we include a plot for $\frac{\mathcal{SC}_{1}(||\mathcal{S}_{p}^{r}||)}{(D-1)}$ and $\frac{\mathcal{SC}_{2}(||\mathcal{S}_{p}^{r}||)}{(D-1)}$ as a function of $\frac{r}{R}$ in Fig. \ref{fig:gsc2S}.

One might be interested in which $\mathcal{SC}_{i}(||\mathcal{S}_{p}^{r}||)$ propagates smaller errors to the estimations of the effective radius $R$ obtained from global fittings or quadratic approximations. This can be done by computing \begin{equation}
    (\frac{\delta R}{R})_{i}:=\delta \mathcal{SC}_{i}(||\mathcal{S}_{p}^{r}||)(R\,\frac{\partial \mathcal{SC}_{i}(||\mathcal{S}_{p}^{r}||)}{\partial R})^{-1},
\end{equation}
where one can notice that using a quadratic approximation of each $\mathcal{SC}_{i}$ one gets $(\frac{\delta R}{R})_{i}<0.03$ for $r<\frac{3}{4}R$, in both $\mathcal{SC}_{i}$. Furthermore, one can see that for fixed $\delta \mathcal{SC}_{i}(||\mathcal{S}_{p}^{r}||)$, $(\frac{\delta R}{R})_{2}<(\frac{\delta R}{R})_{1}$, showing that $\mathcal{SC}_{2}(||\mathcal{S}_{p}^{r}||)$ is likely to provide more robust predictions, as was seen in our results. 

\printbibliography

\end{document}